\documentclass[review,onefignum,oneabnum]{siamart171218}
\usepackage{hyperref}

\usepackage{amsmath,bm,enumitem,amsfonts}									
\usepackage{float}
\usepackage{graphicx,epstopdf}
\usepackage{verbatim}
\usepackage{subfigure}
\usepackage{color}

\newcommand{\TheAuthors}{Luchan Zhang, Xiaoxue Qin and Yang Xiang}

\headers{A unified variational model for grain boundary dynamics}{\TheAuthors}

\title{{A unified variational model for grain boundary dynamics incorporating microscopic structure}\thanks{Submitted to the editors DATE. \funding{This work was  supported by the Hong Kong Research Grants Council General Research Fund 16301720 (Y. Xiang, L.C. Zhang) and Collaborative Research Fund C1005-19G (Y. Xiang),
National Natural Science Foundation of China 12201423 and
Shenzhen Science $\&$ Technology Innovation Program RCYX20231211090222026 (L.C. Zhang),
and Shanghai Sailing Program, China 24YF2712700 (X.X. Qin).}}}

\author{Luchan Zhang\thanks{School of Mathematical Sciences, Shenzhen University, Shenzhen 518060, China
		(\email{zhanglc@szu.edu.cn}).}
\and Xiaoxue Qin\thanks{Department of Mathematics, Shanghai University, Shanghai 200444, China, and Newtouch Center for Mathematics of Shanghai University, Shanghai 200444, China (\email{qinxiaoxue@shu.edu.cn}).}
	\and Yang Xiang\thanks{Department of Mathematics, Hong Kong University of Science and Technology, Clearwater Bay, Kowloon, Hong Kong, and HKUST Shenzhen-Hong Kong Collaborative Innovation Research Institute, Futian, Shenzhen, China (\email{maxiang@ust.hk}).}}
\begin{document}
\nolinenumbers

\maketitle
\begin{abstract}
Recent experiments, atomistic simulations, and theoretical predictions have identified various new types of  grain boundary motions that are controlled by the dynamics of  underlying microstructure of line defects (dislocations or disconnections), to which the classical motion by mean curvature model does not apply. Different continuum models have been developed by upscaling from discrete line defect dynamics models under different settings (dislocations or disconnections, low angle grain boundaries or high angle grain boundaries, etc.), to account for the specific detailed natures of the microscopic dynamics mechanisms, and these continuum models are not in the variational form.
In this paper, we propose a unified variational framework to account for all the underlying line defect mechanisms for the dynamics of both low and high angle grain boundaries and the associated grain rotations.   The variational formulation is based on the developed  constraints of the dynamic Frank-Bilby equations that govern the microscopic line defect structures.
   The proposed variational framework is able to recover the available models for different motions under different conditions.
The unified variational framework is more efficient to describe the collective behaviors of grain boundary networks at larger length scales. It also provides a mathematically tractable basis for rigorous analysis of these partial differential equation models and for the development of efficient numerical methods.
\end{abstract}



\section{Introduction}
Grain boundaries, the interfaces of grains with different orientations,  are surface defects in polycrystalline materials.
The microstructure of a polycrystalline material can be viewed as a network of grain boundaries.
The dynamic properties of grain boundary networks play essential roles in the mechanical and plastic behaviors of the polycrystalline materials~\cite{Sutton1995}.
Grain boundaries migrate under various driving forces such as the capillarity force, the bulk energy difference, the concentration gradients across the boundary, and the  stress field. 
The equilibrium  structure of a grain boundary is governed by the Frank-Bilby equation \cite{HL,Sutton1995}.

The conventional grain boundary dynamics models are based on motion by mean curvature to reduce the total interface energy, where the misorientation angle (or energy density) of the grain boundary is fixed~\cite{Mullins1956,Sutton1995}.
These models can be obtained under the variational framework, i.e., the dynamics is given by gradient flow of the total grain boundary energy with an energy density that does not evolve during the dynamics
\cite{Chenlq1994,Kazaryan2000,liuchun2001,kinderlehrer2006,Selim2009,Srolovitz2010,dai2018convergence,Du-Feng2020}.
Under the variational framework, evolution of misorientation angle was further included in some grain boundary dynamics models to reduce the grain boundary energy density.
In this case, the two grains on different sides of the grain boundary will rotate and cause a relatively rigid-body translation of the two grains along the boundary.
This process is the sliding motion of grain boundaries \cite{Li1962,Shewmon1966,Kobayashi2000,Upmanyu2006,Selim2016,epshteyn2019motion,epshteyn2019large}.

Recent experiments, atomistic simulations, and theoretical predictions have demonstrated that grain boundary motion are controlled by the dynamics of its underlying microstructure of line defects (dislocations or disconnections), which leads to the coupling motion. The grain boundary normal motion can induce a coupled tangential motion which is proportional to the normal motion, as a result of the geometric constraint that the lattice planes must be continuous across the grain boundary \cite{Li1953223,Cahn2002,Cahn20044887}.
In the coupling motion, the energy density can increase although the total energy is decreasing, which cannot be modeled by the available variational models.
Cahn and Taylor \cite{Cahn20044887} proposed a unified theory for the coupling and sliding motions of the grain boundary, which has been examined and generalized by atomistic simulations and experiments \cite{Cahn2002,Cahn20064953,Molodov20071843,Trautt20122407,Wu2012407,Voorhees2016264,Voorhees2017,
Salvalaglio2018053804}. Especially, the coupling motion of the grain boundary is associated with dislocation conservation, and the sliding motion is associated with dislocation reaction. High-angle grain boundary migration is in general controlled by the motion of disconnections (discrete line defects with both characters of dislocations and steps), which have also been demonstrated as the shear-coupling and grain rotation mechanisms in recent theories, atomistic simulations and experiments \cite{Ashby1972498,King1980335,Hirth19964749,Rajabzadeh20131299,Thomas2017,Han2018386,WangJian-npj,Mao2019,WangJian2020,science2022,
Srolovitz3d2024,3drotation2024,science2024}.

We have developed continuum models for the dynamics of low-angle grain boundaries based on their underlying dislocation structures \cite{Zhang2018157,zhang2019new,Qin2021}.
The continuum models incorporate both the motion of the grain boundary and the evolution of the dislocation structure on the grain boundary, and is able to describe the increase of misorientation angle and energy density due to the coupling motion.
We have also developed continuum models of high-angle grain boundary motion based on the underlying discrete mechanisms of disconnections~\cite{Zhang2017119,Wei2019133,Wei2019,Thomas2019,Zhang-npj2021}.
These models demonstrate the physical mechanisms of grain boundary migration with all the driving forces (due to shear and  stress, energy density jump, capillarity) of the grain boundary, mechanical boundary conditions (dislocation reaction rates at triple junctions) and the temperature effect (nucleation of multiple modes of disconnections).
Continuum models have also been developed recently for the migration of high angle grain boundaries and grain rotation based on disconnections for multi-reference planes with grain rotation \cite{Han2021,Han2024}.

These available continuum models for the dynamics of  low angle grain boundaries based on dislocation structure and that of high angle grain boundaries based on disconnection mechanism are obtained by upscaling from discrete line defect dynamics models of dislocations or disconnections, which have both similarities and differences in terms of structure and dynamic mechanism.
 These continuum models contain extra evolutions of the densities of line defects (dislocations or disconnections), in addition to the dynamics of grain boundaries. These continuum models are not directly in the variational form, although the discrete line defect dynamics models, from which the continuum models are derived, are based on energy variations.

In this paper, we develop a unified variational framework to account for all the underlying line defect mechanisms, for the dynamics of both low and high angle grain boundaries. This variational framework incorporates all kinds of motions, including the coupling motion as well as motion by mean curvature and the sliding motion, 
and different driving forces due to grain boundary energy, the stress, and chemical potential jump across the grain boundary. Our continuum models give the grain rotation formulas for both low angle grain boundaries by motion of dislocations and high angle grain boundaries by motion of disconnections.
 We derive dynamic Frank-Bilby equations from the classic, static Frank-Bilby equations~\cite{HL,Sutton1995} that govern the continuum dislocation distributions of low angle grain boundaries or high angle grain boundaries relative to some reference planes. The dynamic Frank-Bilby equations also incorporates the step character of disconnections. The unified variational framework is based on the Onsager principle~\cite{Onsager,Doi2015} with constraints of the dynamic Frank-Bilby equations.
The unified variational framework will be more efficient to describe the collective behaviors of grain boundary networks at larger length scales. The variational framework will also provide basis for rigorous analysis of these partial differential equation models and for the development of efficient numerical methods.

This paper is organized as follows. In Sec.~\ref{sec:constraints}, setting of the continuum framework is presented, and dynamic Frank-Bilby equations are derived. In Sec.~\ref{sec:var}, we present the variational formulation for the dynamics of low angle grain boundaries accounting for the underlying structure of dislocations, based on the Onsager principle and the dynamic Frank-Bilby equations. In Sec.~\ref{sec:comparison},
available models for different motions of grain boundaries are recovered by  different regimes of the grain boundary mobility and dislocation reaction mobility in the obtained unified continuum framework, and numerical simulations in these different regimes are performed. In corporation of  stress and synthetic force in this variational framework is presented in Sec.~\ref{sec:applied}. In Sec.~\ref{sec:high-angle}, we generalize this variational framework to high angle grain boundaries with disconnection structure. Conclusions are given and further generalizations, e.g., to semicoherent hetero-interfaces, are discussed  in Sec.~\ref{sec:conclustions}.

\section{Microscopic constraints: Static and dynamic Frank-Bilby equations}\label{sec:constraints}

The variational grain boundary dynamics model  employs the density of the line defects (dislocations or disconnections) $\mathbf B$ on the grain boundary as variable,  in addition to the grain boundary velocity $\mathbf v$ and misorientation angle $\theta$. This enables the incorporation of various kinds of motions of grain boundaries such as the coupling motion and sliding as well as the classical motion by mean curvature.
Based on the classical Frank-Bilby equations~\cite{HL,Sutton1995} that govern the equilibrium microscopic line defect structure of the grain boundary, we derive
dynamic Frank-Bilby equations which will be used as constraints in our variational formulation.
Here we focus on the low angle grain boundaries which have dislocation structure.
Structure of both dislocations and disconnections of high angle grain boundaries will be considered in Sec.~\ref{sec:high-angle}.

\begin{figure}[htbp]
\centering
    \includegraphics[width=.6\linewidth]{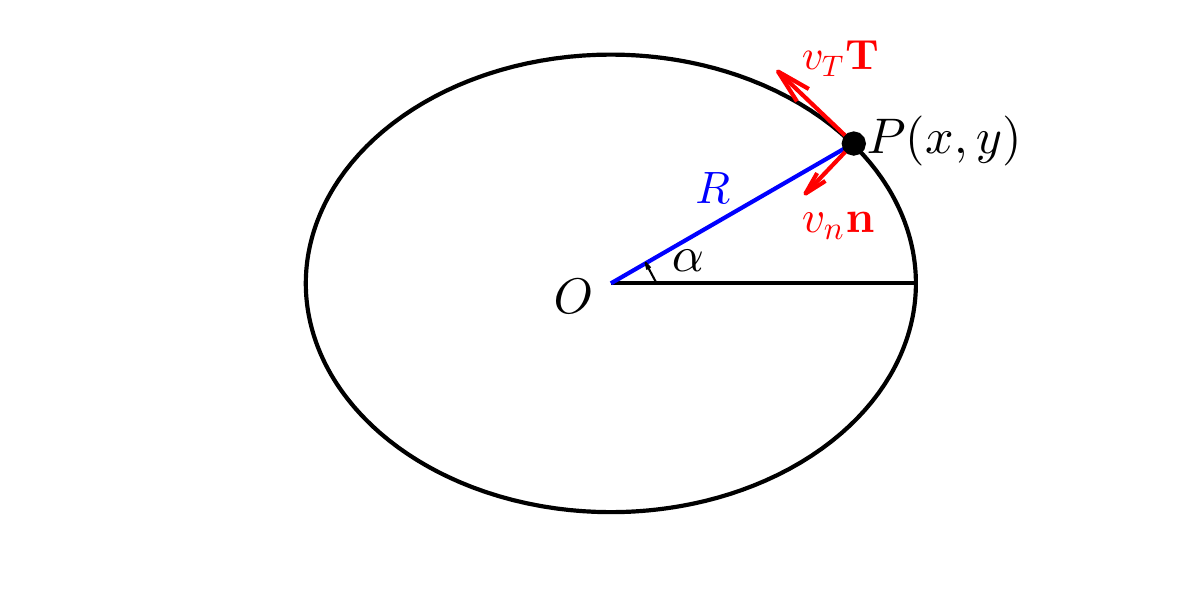}
    \caption{A two dimensional grain boundary which is the cross-section of a cylindrical grain boundary. Its geometric center is $O$. At a point $P(x,y)$ on the grain boundary, the grain boundary velocity is the velocity of the dislocation at this point: $\mathbf v=v_n\mathbf n+v_T\mathbf T$, where  $\mathbf n$ and $\mathbf T$ are the normal and tangent directions of the grain boundary at this point, respectively;  $\alpha$ is the polar angle and $R$ is the radius.  The inner grain has a misorientation angle $\theta$ relative to the outer one.  }
    \label{fig:geometry}
\end{figure}

We focus on the two dimensional setting in this paper. In this case, a grain boundary is a curve in the $xy$ plane (see Fig.~\ref{fig:geometry}), the dislocation density $\mathbf B$ is in the $xy$ plane, and the rotation axis $\mathbf a$ is in the $+z$ direction. For the single grain boundary as shown in Fig.~\ref{fig:geometry}, we set the origin $O$ to be its geometric center, and denote $\mathbf T$ and $\mathbf n$ the tangent and normal directions of the grain boundary, respectively. The inner grain has a misorientation angle $\theta$ relative to the outer one.

 The Frank-Bilby equations~\cite{HL,Sutton1995} govern the equilibrium microscopic line defect structure and provide a link between the macroscopic quantities (including the misorientation angle) and microscopic line defect structure,
which can be written as
$\mathbf B_p=2\sin\frac{\theta}{2}(\mathbf p\times \mathbf a)$,
where  $\theta$ is the misorientation angle, $\mathbf a$ is the rotation axis, and
the dislocation content $\mathbf B_p$ is the closure failure crossing a vector $\mathbf p$ lying in the grain boundary.
The Frank-Bilby equations are equivalent to cancelation of the long-range elastic field generated by the dislocation content~\cite{HL,Sutton1995,Zhu2014175}.
Recall that a dislocation is characterized by its Burgers vector $\mathbf b$ that is the atomic closure failure around the dislocation \cite{HL},
 and for a low angle grain boundary, the dislocation content $\mathbf B_p$ is the sum of the Burgers vectors of all dislocations crossing  vector $\mathbf p$ lying in the grain boundary \cite{HL,Sutton1995}.
 
In the two dimensional case,
the \underline{\bf Frank-Bilby equations} can be written as
\begin{eqnarray}\label{eqn:B0}
 \mathbf B=-2\sin\frac{\theta}{2}\mathbf n,
\end{eqnarray}
where $\mathbf B$ is the dislocation density (dislocation content per unit length) along the grain boundary.

The grain boundary energy density is \cite{HL,Sutton1995,Zhang2018157,Zhang2017126}
\begin{equation}\label{eqn:energydensity}
\gamma(\mathbf B)=\gamma_0B(A_0-\log B),
\end{equation}
where $B=|\mathbf B|$ is the length of $\mathbf B$, $\gamma_0$ is an energy constant and $A_0$ is a dimensionless constant.
Note that in the classical grain boundary energy formula \cite{HL,Sutton1995}, $\gamma=\gamma_0\theta(A_0-\log \theta)$.
For a low angle grain boundary with a single array of dislocations \cite{HL,Sutton1995,Zhang2018157,Zhang2017126},
$\gamma_0=\frac{\mu b}{4\pi(1-\nu)}$ and $A_0=\log(b/r_g)$,
where $b$ is the length of the Burgers vector of the dislocations,   $\mu$ is the shear modulus, $\nu$ is the Poisson ratio, and $r_g$ is a dislocation core parameter. The grain boundary energy in Eq.~\eqref{eqn:energydensity} is based on these formulations and the Frank-Bilby equation in \eqref{eqn:B0}.

The grain boundary velocity $\mathbf v$ at a point $P$  is the velocity of the dislocation located at that point, which has both the normal component  $v_n$ meaning the normal velocity of the grain boundary and the tangential component $v_T$ meaning the moving velocity of the dislocation on the grain boundary, i.e., $\mathbf v=v_n\mathbf n+v_T\mathbf T$. The dislocation density $\mathbf B$ on the grain boundary evolves accordingly with this velocity. The dislocations on the grain boundary may also change due to the dislocation reaction rate $\mathbf B^0_t$, which is the change of dislocation density due to dislocation reaction (nucleation or annihilation of the dislocation content) per unit length  on the grain boundary per unit time.

 We have the following evolutions of geometric quantities
 \begin{flalign}
 \dot{\mathbf n}&=-\left(\frac{dv_n}{ds}+\kappa v_T\right)\mathbf T=-\left(\frac{d\mathbf v}{ds}\cdot\mathbf n\right)\mathbf T,   \label{eqn:ds} \\ \dot{ds}&=\left(-\kappa v_n+\frac{d v_T}{ds}\right)ds=\left(\frac{d\mathbf v}{ds}\cdot\mathbf T\right)ds,\label{eqn:dss}
 \end{flalign}
 where $s$ is the arclength parameter, and $\kappa$ is the curvature of the grain boundary. Recall that  $\frac{d\mathbf T}{ds}=\kappa\mathbf n$ and
$\frac{d\mathbf n}{ds}=-\kappa\mathbf T$.

 The dynamics of dislocation density  $\mathbf B$ is given by
 \begin{flalign}
 \dot{\mathbf B}&
 =-\left(\frac{d\mathbf v}{ds}\cdot\mathbf T\right)\mathbf B +\mathbf B^0_t.\label{eqn:B0t0}
\end{flalign}
Here the dot-notation means time derivative. Note that the first term in this equation is the change of dislocation density $\mathbf B$  due to the change of arclength of the grain boundary during the evolution,
 and the second term $\mathbf B^0_t$ is the dislocation reaction rate.
We have used Eq.~\eqref{eqn:dss}
in the derivation of Eq.~\eqref{eqn:B0t0}.

In the variational formulation, we also need a dynamic form of the Frank-Bilby equations. Note that each of the dislocation density $\mathbf B$, misorientation angle $\theta$ and  the grain boundary normal direction $\mathbf n$ has its own dynamics, and their dynamics
 are related due to the Frank-Bilby equations in Eq.~\eqref{eqn:B0},  leading to non-trivial dynamic Frank-Bilby equations.

The \underline{\bf dynamic Frank-Bilby equations}  are
\begin{eqnarray}\label{eqn:FBderivative18}
\mathbf h=\dot{\theta}\cos\frac{\theta}{2}\mathbf n-2\sin\frac{\theta}{2}\frac{d}{ds}
\left(\mathbf v\times \hat{\mathbf z}\right)+\mathbf B^0_t=\mathbf 0,
\end{eqnarray}
where $\hat{\mathbf z}$ is the unit normal vector in the $+z$ direction (i.e., normal to the $xy$ plane). These dynamic Frank-Bilby equations are derived from the (static) Frank-Bilby equations \eqref{eqn:B0}.
In fact, taking time derivative in \eqref{eqn:B0} and using \eqref{eqn:B0t0} and \eqref{eqn:ds}, we have
$$\dot{\theta}\cos\frac{\theta}{2}\mathbf n-2\sin\frac{\theta}{2}\left(\frac{d\mathbf v}{ds}\cdot\mathbf n\right)\mathbf T  -\left(\frac{d\mathbf v}{ds}\cdot\mathbf T\right)\mathbf B+\mathbf B^0_t=\mathbf 0.$$
Further using the Frank-Bilby equations \eqref{eqn:B0} and $\frac{d\mathbf v}{ds}=\left(\frac{d\mathbf v}{ds}\cdot\mathbf n\right)\mathbf n+\left(\frac{d\mathbf v}{ds}\cdot\mathbf T\right)\mathbf T$, we have \eqref{eqn:FBderivative18}.

The dynamic Frank-Bilby equations in \eqref{eqn:FBderivative18} tells us that the grain rotation, i.e., evolution of misorientation angle $\dot{\theta}$ may be induced either by nucleation of dislocations, or by evolution of the grain boundary. The former is obvious from the classic Frank-Bilby equation. The latter is due to the geometric constraint of the evolution of the dislocation density associated with evolution of the grain boundary; cf. the first contribution
$-\left(\frac{d\mathbf v}{ds}\cdot\mathbf T\right)\mathbf B$
in $\dot{\mathbf B}$ in Eq.~\eqref{eqn:B0t0}.

\underline{Tangential relative velocity and grain rotation}
As reviewed in the introduction, an important property of grain boundary motion not included in the classical motion by mean curvature models is the shear coupling effect, i.e., there is a tangential relative velocity induced by the normal motion of the grain boundary \cite{Cahn20044887,Trautt20122407,Taylor2007493,Zhang2017119,Thomas2017,Zhang2018157,Han2018386,Wei2019133}.  The microscopic mechanism of the shear coupling effect is the plastic shear flow induced by the motion of the constituent  dislocations of the grain boundary.
The shear coupling effect may induce grain rotation \cite{Cahn2002,Cahn20064953,Molodov20071843,Trautt20122407,Wu2012407,Voorhees2016264,Voorhees2017,
Salvalaglio2018053804,Zhang2018157}.

From the obtained dynamic Frank-Bilby equations \eqref{eqn:FBderivative18}, we can derive that
the tangential relative velocity between two grains is
\begin{equation}\label{eqn:parallel}
v_\parallel= 2\tan\frac{\theta}{2} v_n-\frac{1}{\cos\frac{\theta}{2}}\mathbf S\cdot\mathbf T,
\end{equation}
where
\begin{equation}\label{eqn:Cs}
 \mathbf S=\int_0^{s} \mathbf B^0_t(w)dw.
\end{equation}
This is consistent with the Cahn-Taylor theory of the coupling motion \cite{Cahn20044887,Trautt20122407,Taylor2007493,Zhang2018157} with coupling factor $\beta=2\tan\frac{\theta}{2}$. Especially, we can see that following the above discussion on the dynamic Frank-Bilby equations, the coupling effect is due to the geometric constraint of the evolution of the dislocation density associated with evolution of the grain boundary.

The tangential relative velocity $v_\parallel$ is related to grain rotation $\dot{\theta}$   by \cite{Taylor2007493,Zhang2018157}
\begin{equation}\label{eqn:parallel00}
v_\parallel =(R\cos \alpha) \dot{\theta}=(-\mathbf r\cdot \mathbf n)\dot{\theta},
\end{equation}
where $\alpha$ is the
polar angle and $R$ is the radius; see Fig.~\ref{fig:geometry}.
The formulation of tangential relative velocity \eqref{eqn:parallel} can be obtained from the dynamic Frank-Bilby equations \eqref{eqn:FBderivative18}.
In fact,
taking integration of \eqref{eqn:FBderivative18} with respect to $s$, we have another form of the dynamic Frank-Bilby equations
\begin{equation}\label{eqn:h1}
\dot{\theta}\cos\frac{\theta}{2}\mathbf r+2\sin\frac{\theta}{2}\mathbf v+\mathbf S\times\hat{\mathbf z}=\mathbf 0.
\end{equation}
Further using Eq.~\eqref{eqn:parallel00}, we have Eq.~\eqref{eqn:parallel}.
Note that it can be seen from Eq.~\eqref{eqn:FBderivative18} that a necessary condition for this single grain boundary to satisfy the Frank-Bilby equations during the evolution is that
\begin{equation}\label{eqn:bzeroaverage}
\int_\Gamma \mathbf B^0_tds=\mathbf 0.
\end{equation}
This guarantees that $\mathbf S$ in Eq.~\eqref{eqn:Cs} is well-defined.

\section{Variational formulation for grain boundary dynamics}\label{sec:var}

Our unified variational formulation for grain boundary dynamics consists of evolutions of the grain boundary, the dislocation density  $\mathbf B$ on the grain boundary, and the misorientation angle $\theta$.
The formulation is based on the Onsager principle~\cite{Onsager,Doi2015} and constrained minimization subject to the dynamic Frank-Bilby equations \eqref{eqn:FBderivative18}. The independent variables are the grain boundary velocity $\mathbf v$ (i.e., the dislocation velocity), the dislocation reaction rate $\mathbf B^0_t$, and the evolution of the misorientation angle $\dot{\theta}$, together with the dynamics of dislocation density  $\mathbf B$ in \eqref{eqn:B0t0} and the Frank-Bilby equations \eqref{eqn:B0}.

The constrained minimization problem is
\begin{eqnarray}
{\rm minimize \ } &&
 Q+\dot{E} \vspace{1ex}\label{eqn:onsager}\\
{\rm subject \ to \ }&&  \mathbf h=\mathbf 0. \label{eqn:onsager-h}
\end{eqnarray}
This minimization problem determines the grain  boundary velocity  $\mathbf v$, the reaction rate of dislocation density  $\mathbf B_t^0$, and evolution of misorientation angle $\dot{\theta}$. The constraint $\mathbf h=\mathbf 0$   is the dynamic Frank-Bilby equations in \eqref{eqn:FBderivative18}, and is  based on the assumption that the grain boundary velocity, the dislocation reaction rate, and the evolution of the misorientation angle are adjusted to satisfy the Frank-Bilby equations over a timescale that is much shorter than that of the evolutions of the grain boundary and the dislocation density.

In the minimization problem,
$Q$ is the dissipation function:
\begin{flalign}\label{eqn:dissipation}
Q=\int_\Gamma \left(\frac{1}{2M_n}\mathbf v^2+\frac{1}{2M_B}{\mathbf B^0_t}+\frac{1}{2M_\theta}\dot{\theta}^2
\right) ds,
\end{flalign}
where recall that $\mathbf v=v_n\mathbf n+v_T\mathbf T$ is the grain boundary velocity,  $\mathbf B^0_t$ is the rate of change of dislocation density due to dislocation reaction, $\dot{\theta}$ is the evolution of the misorientation angle, and   $M_n$, $M_B$ and $M_\theta$ are mobilities. Here we assume that $v_n$ and $v_T$ have the same mobility $M_n$. This assumption can be easily relaxed.
The term
$\dot{E}$ is the dissipation rate of the total energy
\begin{equation}\label{eqn:energy}
E=\int_\Gamma \gamma(\mathbf B)ds,
\end{equation}
where $\gamma(\mathbf B)$ is the grain boundary energy density depending on the dislocation density $\mathbf B$.
The rate of energy dissipation is
\begin{flalign}\label{eqn:energy-dissipation}
 \dot{E}=&\int_\Gamma\left[\gamma(\mathbf B)\left(\frac{d\mathbf v}{ds}\cdot\mathbf T\right)+\frac{\partial \gamma}{\partial \mathbf B}\cdot\left( -\left(\frac{d\mathbf v}{ds}\cdot\mathbf T\right)\mathbf B+\mathbf B^0_t\right) \right]ds.
\end{flalign}
Here we have used the expressions of $\dot{ds}$ and $\dot{\mathbf B}$ in Eqs.~\eqref{eqn:dss} and \eqref{eqn:B0t0}.

Now we solve the constrained minimization problem of Eqs.~\eqref{eqn:onsager} and \eqref{eqn:onsager-h}.
The Lagrangian function of the minimization problem with the constraint $\mathbf h=0$ is
\begin{flalign}\label{eqn:Lagrange}
L=&Q+\dot{E}+\int_\Gamma \pmb\lambda \cdot \mathbf h ds.
\end{flalign}
Taking variations of the Lagrangian function, we have
\begin{flalign}
\frac{\delta L}{\delta \mathbf v}=&\frac{1}{M_n}\mathbf v-\frac{1}{M_n}\mathbf v^*-2\sin\dfrac{\theta}{2}\frac{d}{d s}\left(\pmb\lambda\times \hat{\mathbf z}\right)=\mathbf 0,\label{eqn:l1}\\
\frac{\delta L}{\delta \mathbf B^0_t}=& \frac{1}{M_B}\mathbf B^0_t-\frac{1}{M_B}\mathbf {B^{0*}_t}+\pmb \lambda=\mathbf 0,\label{eqn:l3}\\
\frac{\delta L}{\delta \dot{\theta}}=&\frac{l}{M_\theta} \dot{\theta}+\cos\dfrac{\theta}{2}\int_\Gamma\pmb\lambda\cdot\mathbf n ds=0,\label{eqn:l4}
\end{flalign}
where $l$ is the perimeter of the grain boundary $\Gamma$, $\mathbf v^*$ and $\mathbf B_t^{0*}$ are
the grain boundary velocity  and rate of change of dislocation density due to dislocation reaction, respectively, obtained in the unconstrained minimization problem $\min \{L^*=Q+\dot{E}\}$, i.e., $\frac{\delta L^*}{\delta \mathbf v}=\mathbf 0$ and $\frac{\delta L^*}{\delta \mathbf B_t^0}= \mathbf 0$, and the formulations are
\begin{flalign}
\mathbf v^*=&-M_n\frac{\delta E}{\delta \mathbf v}=M_n\frac{d}{ds}\left[\left(\gamma(\mathbf B)-\frac{\partial \gamma}{\partial \mathbf B}\cdot \mathbf B\right)\mathbf T\right],\label{eqn:v*}\\
\mathbf {B^0_t}^*=&-M_B\frac{\delta E}{\delta \mathbf B}=-M_B\dfrac{\partial \gamma}{\partial \mathbf B}.\label{eqn:B*}
\end{flalign}

We solve for $\mathbf v$, $\mathbf B_t^0$,  $\dot{\theta}$ and $\pmb\lambda$ from Eqs.~\eqref{eqn:l1}--\eqref{eqn:l4} together with the constraint \eqref{eqn:onsager-h}. 
From Eqs.~\eqref{eqn:l1} and \eqref{eqn:l3}, we have
\begin{flalign}
\frac{1}{M_n}(\mathbf v-\mathbf v^*)+\frac{2\sin\frac{\theta}{2}}{M_B}\frac{d}{d s}\big((\mathbf B^0_t-\mathbf {B^{0*}_t})\times\hat{\mathbf z}\big)=\mathbf 0.
\label{eqn:l3c2}
\end{flalign}
Note that this gives
\begin{equation}
\int_\Gamma  \mathbf v ds=\int_\Gamma  \mathbf v^* ds=\mathbf 0. \label{eqn:full-cond}\\
\end{equation}

From Eq.~\eqref{eqn:l3c2}  and the dynamic Frank-Bilby equations \eqref{eqn:FBderivative18} (i.e. \eqref{eqn:onsager-h}), it can be solved  using Eqs.~\eqref{eqn:l3} and \eqref{eqn:l4}, that $\mathbf v$ and $\dot{\theta}$  satisfy
\begin{flalign}
 &-\left(2\sin\frac{\theta}{2}\right)^2\frac{M_n}{M_B}\frac{d^2 \mathbf v }{ds^2}+\mathbf v=\mathbf{f_v},\label{odev}\\
&\dot{\theta}=-M_\theta \frac{\cos\dfrac{\theta}{2}}{l}\int_\Gamma\pmb\lambda\cdot\mathbf n ds,\label{eqn:ll4}
\end{flalign}
where $\pmb\lambda$, $\mathbf{f_v}$ and $\mathbf{f_{\pmb\lambda}}$ are given by
\begin{flalign}
 &-\left(2\sin\frac{\theta}{2}\right)^2\frac{M_n}{M_B}\frac{d^2 \pmb\lambda}{ds^2}+\pmb\lambda=\mathbf{f_{\pmb\lambda}},\label{odelambda}\\
&\mathbf{f_v}=\frac{2\sin\frac{\theta}{2}M_n}{M_B}\dot{\theta}\cos\frac{\theta}{2}\kappa \mathbf n
+\mathbf v^*+\frac{2\sin\frac{\theta}{2}M_n}{M_B}\frac{d}{ds}\left(\mathbf {B^0_t}^*\times\hat{\mathbf z}\right),\label{eqn:fv}\\
&\mathbf{f_{\pmb\lambda}}=\frac{1}{M_B}\left[\dot{\theta}\cos\frac{\theta}{2}\mathbf n -2\sin\frac{\theta}{2}\frac{d}{ds}
\left(\mathbf v^*\times \hat{\mathbf z}\right)+{\mathbf B^0_t}^*\right]. \label{eqn:flambda}
\end{flalign}
Note that \eqref{eqn:ll4} is an equation for $\dot{\theta}$ because $\pmb\lambda$ depends on $\dot{\theta}$ linearly. The  dislocation density $\mathbf B$  during the evolution can be calculated by the static Frank-Bilby equations \eqref{eqn:B0} associated with the grain boundary profile. 

In summary, the variational formulation for grain boundary dynamics consists of Eqs.~\eqref{eqn:onsager} and \eqref{eqn:onsager-h}, with Eqs.~\eqref{eqn:dissipation}--\eqref{eqn:energy-dissipation}, \eqref{eqn:B0t0} and the static Frank-Bilby equations \eqref{eqn:B0}, to determine the
the grain boundary velocity $\mathbf v$, 
the dislocation density reaction rate $\mathbf B^0_t$,
and evolution of the misorientation angle $\dot{\theta}$.
The resulting grain boundary dynamics equations are
Eqs.~\eqref{odev}--\eqref{eqn:flambda}, 
and \eqref{eqn:B0}.
The coupling tangential relative velocity is given by Eq.~\eqref{eqn:parallel}.

Sometimes, we may also need $\mathbf B_t^0$ during the evolution, which can be obtained by solving a similar ODE
\begin{flalign}
 &-\left(2\sin\frac{\theta}{2}\right)^2\frac{M_n}{M_B}\frac{d^2 \mathbf B^0_t}{ds^2}+\mathbf B^0_t=\mathbf{f_B},\label{odeB}\\
&\mathbf{f_B}=-\dot{\theta}\cos\frac{\theta}{2}\mathbf n+2\sin\frac{\theta}{2}\frac{d}{ds}\left(\mathbf v^*\times\hat{\mathbf z}\right)
-\frac{\left(2\sin\frac{\theta}{2}\right)^2M_n}{M_B}\frac{d^2 \mathbf {B^0_t}^*}{ds^2}. \label{eqn:fB}
\end{flalign}

This variational formulation provides a unified framework for the classical motion by mean curvature, the sliding motion, and the coupling motion as reviewed in the introduction. Different types of grain boundary motions can be recovered by different regimes of the grain boundary mobility $M_n$, the dislocation reaction mobility $M_B$, and the grain rotation mobility $M_\theta$.
These will be discussed in details in the next section.

For the numerical solution, from Eq.~\eqref{eqn:ll4}, we have
\begin{flalign}
\dot{\theta}=&-\frac{M_\theta}{M_B+M_\theta\frac{\cos^2\frac{\theta}{2}}{l}\int_\Gamma \mathbf p^{n}\cdot\mathbf n ds}\frac{\cos\frac{\theta}{2}}{l}\int_\Gamma \mathbf p^*\cdot\mathbf n ds.\label{eqn:ll5}
\end{flalign}
Here we have used the solution formula $\pmb\lambda=\frac{1}{M_B}\dot{\theta}\cos\frac{\theta}{2}\mathbf p^n+\frac{1}{M_B}\mathbf p^*$ from Eqs.~\eqref{odelambda} and \eqref{eqn:flambda}, where $\mathbf p^n$ and $\mathbf p^*$ are the solutions of the ODEs $-\left(2\sin\frac{\theta}{2}\right)^2\frac{M_n}{M_B}\frac{d^2 \mathbf u }{ds^2}+\mathbf u=\mathbf{f}$ with $\mathbf{f}=\mathbf n$ and
$ \mathbf f=-2\sin\frac{\theta}{2}\frac{d}{ds}\left(\mathbf v^*\times \hat{\mathbf z}\right)+{\mathbf B^0_t}^*$, respectively. See Appendix~\ref{appendix A} for the solution formulas of these ODEs.
With the solved  $\dot{\theta}$ in Eq.~\eqref{eqn:ll5}, solution formulas of $\mathbf v$ can also be obtained by Eqs.~\eqref{odev} and \eqref{eqn:fv}  using the ODE solution formula given in Appendix~\ref{appendix A}.
As described above, the  dislocation density $\mathbf B$  during the evolution can be calculated by the static Frank-Bilby equations \eqref{eqn:B0} associated with the grain boundary profile. 

\section{Comparisons with available models for different motions of grain boundaries}\label{sec:comparison}

In this section, we show that our unified variational grain boundary dynamics model is able to recover the classical and recently developed models for different motions of grain boundaries, in different regimes of the grain boundary mobility $M_n$, the dislocation reaction mobility $M_B$, and the grain rotation mobility $M_\theta$.

\subsection{Slow dislocation reaction: Coupling motion}\label{subsec:slow}
We first consider the case of slow dislocation reaction, i.e., $M_n/ M_B\gg l^2/B^2$, where $l$ is the length scale of the continuum model, and $B=|\mathbf B|$.
Note that using the  Frank-Bilby equations \eqref{eqn:B0}, $B=2\left|\sin\frac{\theta}{2}\right|$.
In this case, the mobility $M_n$ for the grain boundary velocity is much larger than the mobility $M_B$ for the reaction of dislocations, and the dynamic Frank-Bilby equations \eqref{eqn:FBderivative18} are maintained mainly by adjustment of the grain boundary velocity $\mathbf v$, and the grain rotation $\dot{\theta}$ mainly comes from the unconstrained variational velocity $\mathbf v^*$.

In this limit, in  Eq.~\eqref{odeB}, the terms $\mathbf B^0_t$ and $\dot{\theta}\cos\frac{\theta}{2}\mathbf n$
can be dropped compared with $\left(2\sin\frac{\theta}{2}\right)^2\frac{M_n}{M_B}\frac{d^2 \mathbf B^0_t}{ds^2}$, and further integrating the equation twice, we have
\begin{flalign}
 \mathbf B^0_t=
  M_B\left[-\frac{\partial \gamma}{\partial \mathbf B}+\frac{1}{2\sin\frac{\theta}{2}}
\left(\gamma(\mathbf B)-\frac{\partial \gamma }{\partial \mathbf B}\cdot \mathbf B\right)\mathbf n\right]+\mathbf C_B
=\frac{M_B}{2\sin\frac{\theta}{2}}(\gamma\mathbf n-\gamma' \mathbf T)+\mathbf C_B.
\label{eqn:l3c3}
\end{flalign}
Here the constant $\mathbf C_B$ is a constant such that Eq.~\eqref{eqn:bzeroaverage} holds, $\gamma'=\frac{d\gamma}{d\alpha}$ where $\alpha$ is the inclination angle (c.f. Fig.~\ref{fig:geometry}), and the last equation in \eqref{eqn:l3c3} is obtained using the Frank-Bilby equation \eqref{eqn:B0} and the relations $\mathbf n=(\cos\alpha,\sin\alpha)$ and $\frac{d\gamma}{d\alpha}=-\frac{\partial \gamma}{\partial \mathbf n}\cdot\mathbf T$.

In Eq.~\eqref{odev}, in this slow dislocation reaction regime, the term $\mathbf v$ can be dropped compared with $\left(2\sin\frac{\theta}{2}\right)^2\frac{M_n}{M_B}\frac{d^2 \mathbf v }{ds^2}$. Integrating it twice and using Eq.~\eqref{eqn:l3c3}, we have
the dynamic Frank-Bilby equations \eqref{eqn:h1}, which give
 \begin{flalign}\label{eqn:v-slow}
\mathbf v=-\frac{\dot{\theta}\cos\frac{\theta}{2}}{2\sin \frac{\theta}{2}}\mathbf r-\frac{1}{2\sin\frac{\theta}{2}}\mathbf S\times \hat{\mathbf z},
\end{flalign}
where $S$ is given by Eq.~\eqref{eqn:Cs}.

Similarly, in Eq.~\eqref{odelambda}, in this slow dislocation reaction regime, the term $\pmb\lambda$ can be dropped compared with $\left(2\sin\frac{\theta}{2}\right)^2\frac{M_n}{M_B}\frac{d^2 \pmb\lambda }{ds^2}$. Thus we have $\frac{d^2 \mathbf p^n}{ds^2}=-\frac{M_B}{\left(2\sin\frac{\theta}{2}\right)^2M_n}\mathbf n$ and
$\frac{d^2 \mathbf p^*}{ds^2}=-\frac{M_B}{\left(2\sin\frac{\theta}{2}\right)^2M_n}\left[ -2\sin\frac{\theta}{2}\frac{d}{ds}\left(\mathbf v^*\times \hat{\mathbf z}\right)+{\mathbf B^0_t}^*\right]$. Following Eq.~\eqref{eqn:ll5}, the evolution of the misorientation angle in this case is
\begin{flalign}
 \dot{\theta}=&-\frac{M_\theta \sin\theta }{\left(2\sin\frac{\theta}{2}\right)^2M_n+M_\theta\cos^2\frac{\theta}{2}\frac{1}{l}\int_\Gamma r^2ds}\frac{1}{l}\int_\Gamma \mathbf v^*\cdot\mathbf r \ ds.\label{eqn:ll6}
\end{flalign}

In this case, the grain boundary dynamics equations consist of the velocity $\mathbf v$ in Eq.~\eqref{eqn:v-slow}, the dislocation reaction rate $\mathbf B^0_t$ in Eq.~\eqref{eqn:l3c3}, and the evolution of misorientation angle $\dot{\theta}$ in Eq.~\eqref{eqn:ll6}.   
The coupled tangential relative velocity is given by Eq.~\eqref{eqn:parallel}.

For the pure coupling motion, there is no dislocation reaction: $M_B=0$ and $\mathbf B^0_t=0$.
In this case, $\mathbf S=\mathbf 0$. The grain boundary dynamic model 
is reduced to
\begin{flalign}
&\mathbf v=-\frac{\dot{\theta}\cos\frac{\theta}{2}}{2\sin \frac{\theta}{2}}\mathbf r.
\end{flalign}

These obtained formulations 
in this slow dislocation reaction regime recover the previously proposed continuum model for the grain boundary dynamics incorporating both  effects of the coupling and sliding motions \cite{Zhang2018157,zhang2019new}.
Especially, the velocity is in the inward radial direction of the grain boundary and the shape of grain boundary is preserved. This is in the same form as the continuum model derived in Ref.~\cite{zhang2019new} for low angle grain boundaries.
This grain boundary dynamics under combined effect of coupling motion under conservation of total dislocation numbers and sliding motion due to dislocation reactions
agrees with the available theoretical analyses \cite{Cahn20044887,Taylor2007493,Zhang2018157}, continuum simulations \cite{Zhang2018157}, atomistic simulations
\cite{Cahn2002,Trautt20122407,Wu2012407} and discrete dislocation dynamics simulations \cite{Zhang2018157}.
The obtain formulation is also consistent with the three dimensional atomistic simulations by phase field crystal model~\cite{Voorhees2017} and by amplitude expansion phase
field crystal model \cite{Salvalaglio2018053804}, and the continuum model and simulation results for grain boundary dynamics in three dimensions that incorporates these two effects \cite{Qin2021}.

\subsection{Fast dislocation reaction: Sliding motion}\label{subsec:fast}

Next, we consider the case of fast dislocation reaction, i.e., $M_n/ M_B\ll l^2/B^2$. In this case,
the mobility for dislocation reaction $M_B$ is much larger than the mobility $M_n$ of grain boundary dynamics,
and the Frank-Bilby equations \eqref{eqn:FBderivative18} are maintained mainly by dislocation reaction, and the grain rotation $\dot{\theta}$ mainly comes from the unconstrained variational dislocation density reaction rate $\mathbf {B_t^0}^*$.

In this limit, in Eq.~\eqref{odev}, the two terms with the factor $M_n/ M_B$ can be dropped, and we have 
\begin{flalign}
\mathbf v= M_n\frac{d}{ds}\left[\left(\gamma(\mathbf B)-\frac{\partial \gamma}{\partial \mathbf B}\cdot \mathbf B\right)\mathbf T-2\sin\frac{\theta}{2}\frac{\partial \gamma}{\partial \mathbf B}\times\hat{\mathbf z}\right]=M_n(\gamma+\gamma'')\kappa \mathbf n. \label{eqn:fast-reaction}
\end{flalign}
Here $\gamma''$ is the second derivative of $\gamma$ with respect to the inclination angle $\alpha$; see Fig.~\ref{fig:geometry}, and the last equation in Eq.~\eqref{eqn:fast-reaction} is obtained using  $-\frac{\partial \gamma}{\partial \mathbf B}+\frac{1}{2\sin\frac{\theta}{2}}
\left(\gamma(\mathbf B)-\frac{\partial \gamma }{\partial \mathbf B}\cdot \mathbf B\right)\mathbf n=\frac{1}{2\sin\frac{\theta}{2}}(\gamma\mathbf n-\gamma' \mathbf T)$ in Eq.~\eqref{eqn:l3c3},
and the relation  $\frac{d}{ds}= \kappa\frac{d}{d\alpha}$. Note that this grain boundary velocity is independent of the dislocation reaction rate $\mathbf B_t^0$.
This velocity is the same as the velocity in the classical motion by mean curvature models \cite{Herring1951,Mullins1956,Sutton1995}. When the misorientation angle is fixed, i.e., $\dot{\theta}=0$ by $M_\theta=0$ in Eq.~\eqref{eqn:ll5}, our variational model recovers these classical motion by mean curvature models.

The dislocation reaction rate $\mathbf B_t^0$ is obtained by Eq.~\eqref{odeB} in which the term with the factor $M_n/ M_B$ is dropped:
\begin{flalign}
\mathbf B_t^0=& -\dot{\theta}\cos\frac{\theta}{2}\mathbf n
+2\sin\frac{\theta}{2}M_n\frac{d^2}{ds^2}\left[2\sin\frac{\theta}{2}\frac{\partial \gamma}{\partial \mathbf B}- \left(\gamma(\mathbf B)-\frac{\partial \gamma }{\partial \mathbf B}\cdot \mathbf B\right)\mathbf n \right]\nonumber\\
=&-\dot{\theta}\cos\frac{\theta}{2}\mathbf n+2\sin\frac{\theta}{2}M_n\frac{d}{ds}\big[(\gamma+\gamma'')\kappa \mathbf T\big].\label{eqn:fast-reactionB}
\end{flalign}
Note that this equation is equivalent to the dynamic Frank-Bilby equations \eqref{eqn:FBderivative18} when the grain boundary velocity $\mathbf v$ is given by Eq.~\eqref{eqn:fast-reaction}.

Similarly, in Eq.~\eqref{odelambda}, in this fast dislocation reaction regime, the term with the factor $M_n/ M_B$ is dropped,
thus we have $\mathbf p^n=\mathbf n$ and
$\mathbf p^*= -2\sin\frac{\theta}{2}\frac{d}{ds}\left(\mathbf v^*\times \hat{\mathbf z}\right)+{\mathbf B^0_t}^*$. Following Eq.~\eqref{eqn:ll5}, the evolution of the misorientation angle in this case is
\begin{flalign}
\dot{\theta}=&-\frac{M_\theta \cos\frac{\theta}{2}}{M_B+M_\theta \cos^2\frac{\theta}{2}}\frac{1}{l}\int_\Gamma{\mathbf B_t^0}^*\cdot\mathbf nds.\label{eqn:ll8}
\end{flalign}
In the limit case of large $M_B$, this evolution of misorientation angle is
\begin{flalign}
\dot{\theta}=&M_\theta \frac{\cos\frac{\theta}{2}}{l}\int_\Gamma\frac{\partial \gamma}{\partial \mathbf B}\cdot\mathbf nds.\label{eqn:ll9}
\end{flalign}
Using Frank-Bilby equations \eqref{eqn:B0}, we have
$\frac{\partial \gamma}{\partial \theta}=\frac{\partial \gamma}{\partial \mathbf B}\cdot
\frac{\partial \mathbf B}{\partial \theta}=-\cos\frac{\theta}{2}\frac{\partial \gamma}{\partial \mathbf B}\cdot\mathbf n$. Thus
\begin{flalign}
 \dot{\theta} =-M_\theta\frac{\partial \gamma}{\partial \theta}.
 \label{eqn:ll10}
\end{flalign}

Eqs.~\eqref{eqn:fast-reaction}  and  \eqref{eqn:ll10} recover the available grain boundary dynamic models with grain boundary sliding \cite{Li1962,Shewmon1966,Kobayashi2000,Upmanyu2006,Selim2016,epshteyn2019motion,epshteyn2019large}.

\subsection{Numerical simulations of different dislocation reaction rates}\label{subsec:simulation}
In this subsection,  we further validate our
unified continuum model of grain boundary motion with dislocation structure in Eqs.~\eqref{odev}, 
\eqref{eqn:fv}, 
\eqref{eqn:ll5}, 
and \eqref{eqn:B0}, by  numerical simulations for different dislocation reaction rates: slow, fast, and median, compared with the grain boundary motion.

In our simulations, we set $M_n=M_d b/B$, $M_B=K_B M_dB/b$, and $M_\theta=K_\theta M_d/b$. Here the mobilities are expressed in terms of the mobility of dislocations $M_d$,  the dislocation number density $B/b$ and the length of the Burgers vector $b$, and $K_B$ and $K_{\theta}$ are dimensionless constants. The formula of $M_n$ is from the discrete dislocation dynamics model \cite{Zhang2018157}. The dependence of $M_B$ on the dislocation number density $B/b$ because the dislocation reaction rate depends on the dislocation number density, which is consistent with the atomistic simulations \cite{Trautt20122407}. The grain boundary energy density $\gamma(\mathbf B)$ is given by Eq.~\eqref{eqn:energydensity}
with $\gamma_0=\frac{\mu b}{4\pi(1-\nu)}$ and $A_0=\log(b/r_g)$.

The initial grain boundary profile is set as an ellipse, with two axes $60b$ and $40b$, respectively.
We choose three values for the dislocation reaction mobility: $M_B=2.86\times 10^{-7}M_dB/b$, $2.86\times10^{-5}M_dB/b$, and $8.58\times 10^{-5}M_dB/b$, and if the length scale of grain boundary perimeter is approximated by $2\pi l\approx314b$, these correspond to the cases of (i) $M_n/ M_B=35.5l^2/B^2$, (ii) $M_n/ M_B=0.355l^2/B^2$, and (iii) $M_n/ M_B=0.118l^2/B^2$, respectively.
The mobility associated with grain rotation $M_\theta=2.34\times10^{-3}M_d/b$.
The time unit is $t_0=3.24\times 10^4/M_d\mu$. The length of the Burgers vector is $b=0.286$nm, Poisson ratio $\nu= 0.347$, the grain boundary core parameter $r_g=b$, as in Al.

\begin{figure}[htbp]
\centering
\subfigure[]    {\includegraphics[width=.45\linewidth]{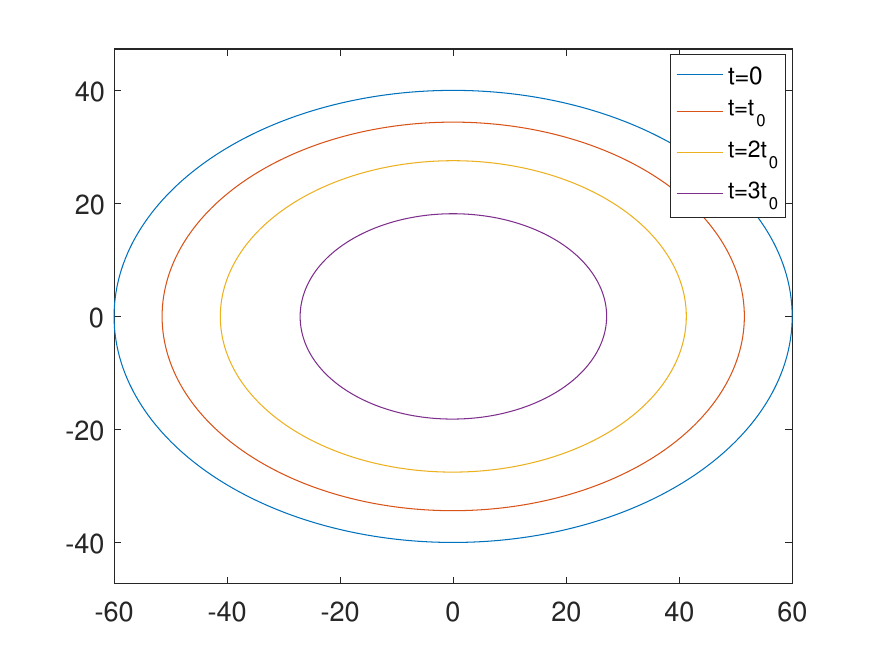}}
\subfigure[]    {\includegraphics[width=.45\linewidth]{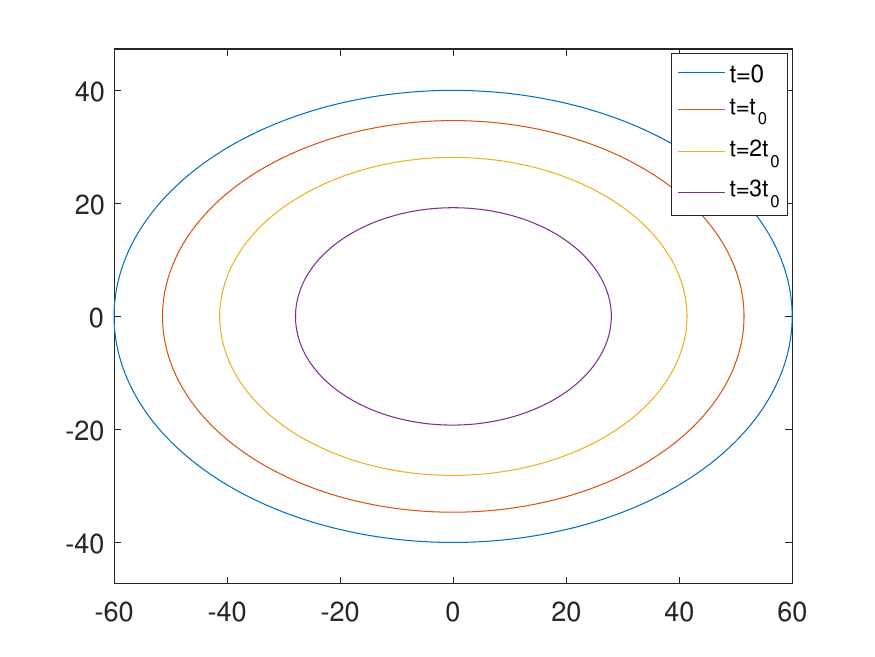}}
\subfigure[]    {\includegraphics[width=.45\linewidth]{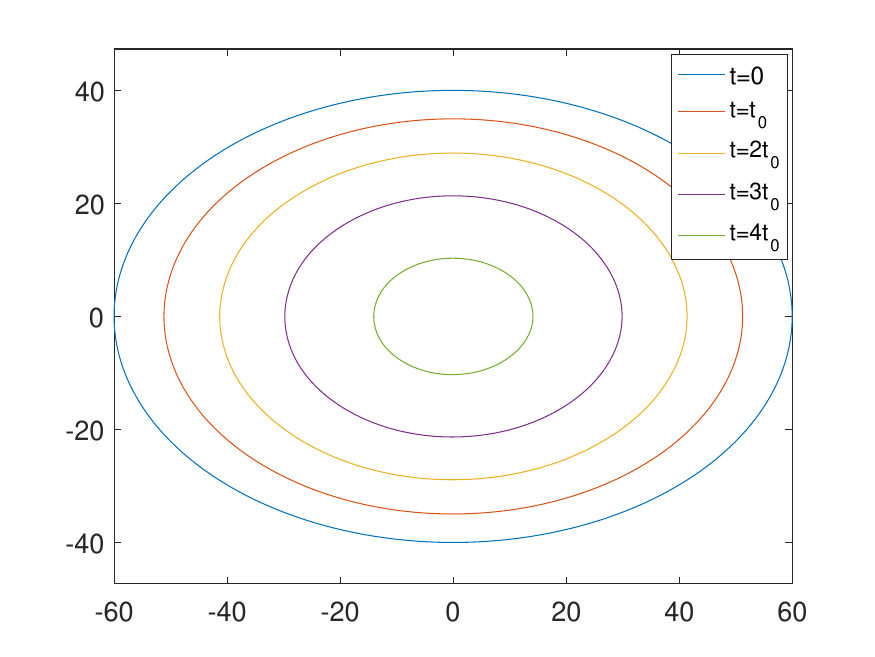}}
\subfigure[]    {\includegraphics[width=.45\linewidth]{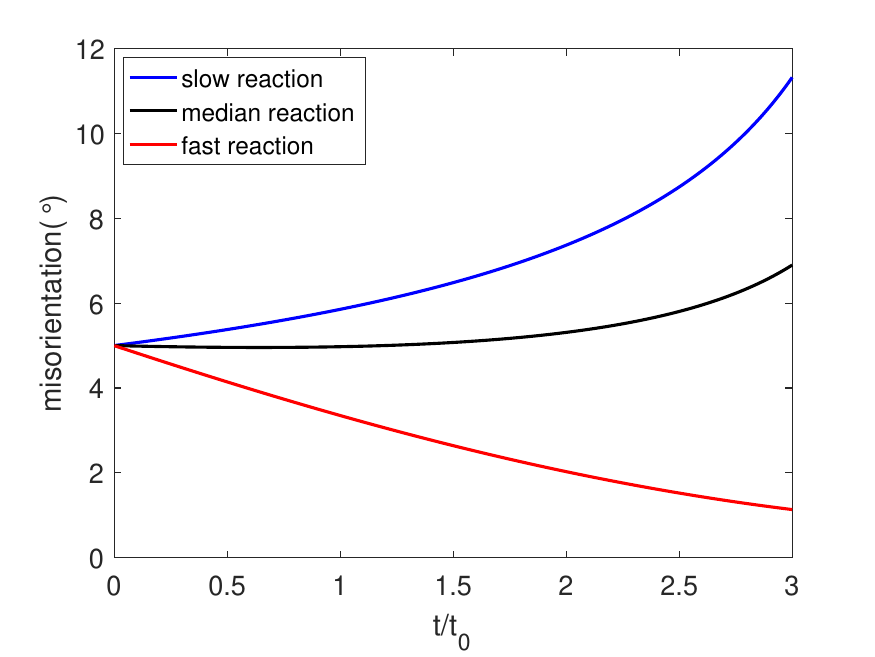}}
    \caption{The grain boundary motion when (a) $M_n/ M_B=35.5l^2/B^2$, (b) $M_n/ M_B=0.355l^2/B^2$,  (c) $M_n/ M_B=0.118l^2/B^2$, and (d) evolution of the misorientation $\theta$.}
    \label{fig:simulation}
\end{figure}

The simulation result of grain boundary evolution of case (i) is shown in Fig.~\ref{fig:simulation}(a). In this case,
 $M_n/ M_B\gg l^2/B^2$, and the dislocations on the grain boundary have slow reactions compared to the grain boundary motion. 
It can be observed that the grain boundary moves in the inward radial direction and the shape of grain boundary is preserved. Especially, in the last grain boundary profile, the ratio of the major axis over the minor axis is about $1.5$, which equals that in the initial profile.  The misorientation angle is increasing  during this evolution process as shown in Fig.~\ref{fig:simulation}(d). Accordingly, the grain boundary energy density is also increasing during the evolution. These are consistent with the theoretical analysis in Sec.~\ref{subsec:slow} as well as the continuum models presented in Ref.~\cite{Zhang2018157,zhang2019new} for the dynamics of low angle grain boundaries in the slow dislocation reaction regime.

The simulation result of grain boundary evolution of case (iii) is shown in Fig.~\ref{fig:simulation}(c). In this case,
 $M_n/ M_B\ll l^2/B^2$, and the dislocations on the grain boundary have fast reaction compared to the grain boundary motion. 
It can be observed that the grain boundary gradually changes towards a circular shape as it shrinks. In the last grain boundary profile, the ratio of the major axis over the minor axis is about $1.35$, which is smaller than the value $1.5$  in the initial profile.
The misorientation angle is decreasing during this evolution process as shown in Fig.~\ref{fig:simulation}(d). Accordingly, the grain boundary energy density is also decreasing during the evolution.
 These are consistent with the theoretical analysis in Sec.~\ref{subsec:fast} as well as the continuum models presented in Ref.~\cite{Zhang2018157,zhang2019new} for the dynamics of low angle grain boundaries in the fast dislocation reaction regime.

The simulation result of grain boundary evolution of case (ii) is shown in Fig.~\ref{fig:simulation}(b)
corresponding to the case $M_n/ M_B\simeq l^2/B^2$. In this case, the reaction of the dislocations on the grain boundary is comparable with the grain boundary motion.
It can be observed from the simulation results that this is an intermediate case between the slow and fast dislocation reaction cases in Fig.~\ref{fig:simulation}(a) and (c). There is a slight  tendency towards  a circular shape during the grain boundary evolution.  In the last grain boundary profile, the ratio of the major axis over the minor axis is about $1.44$, which is slightly smaller than the value $1.5$  in the initial profile.
The grain boundary shrinking and the increase of misorientation angle  (shown in Fig.~\ref{fig:simulation}(d)) in this case are both slow compared to the those of the slow dislocation reaction case.

\section{Effects of stress and synthetic force}\label{sec:applied}

In this section, we include the effects of  stress and synthetic force in the variational model.
With  these effects, the rate of energy dissipation in Eq.~\eqref{eqn:energy-dissipation} is changed to
\begin{flalign}\label{eqn:energy-dissipation1}
 \dot{E}=&\int_\Gamma\left[\gamma(\mathbf B)\left(\frac{d\mathbf v}{ds}\cdot\mathbf T\right)+\frac{\partial \gamma}{\partial \mathbf B}\cdot\left(
 -\left(\frac{d\mathbf v}{ds}\cdot\mathbf T\right)\mathbf B+\mathbf B^0_t\right) \right.\nonumber\\
 & -((\pmb \sigma\cdot\mathbf B)\times \hat{\mathbf z})\cdot \mathbf v+ \Psi (\mathbf v\cdot\mathbf n)\bigg]ds,
\end{flalign}
where $\pmb \sigma$ is the  stress tensor, and $\Psi$ is the synthetic force. The constrained energy minimization problem now consists of
 Eqs.~\eqref{eqn:onsager} and \eqref{eqn:onsager-h}, with Eqs.~\eqref{eqn:dissipation}, \eqref{eqn:energy-dissipation}, \eqref{eqn:B0t0} and the static Frank-Bilby equations \eqref{eqn:B0}.

The Lagrange function of this minimization problem and its variations are still given  in Eq.~\eqref{eqn:Lagrange}--\eqref{eqn:l4}, with
\begin{flalign}
\mathbf v^*=&M_n\left[\frac{d}{ds}\left(\left(\gamma(\mathbf B)-\frac{\partial \gamma}{\partial \mathbf B}\cdot \mathbf B\right)\mathbf T\right) +(\pmb \sigma\cdot\mathbf B)\times \hat{\mathbf z} -\Psi\mathbf n\right],\label{eqn:v-star-stress}
\end{flalign}
and same $\mathbf B^{0*}_t$ in Eq.~\eqref{eqn:B*}.

The resulting grain boundary dynamics equations are still
 Eqs.~\eqref{odev},
\eqref{eqn:fv},
\eqref{eqn:ll5},
and \eqref{eqn:B0},
in which $\mathbf v^*$ is given by Eq.~\eqref{eqn:v-star-stress}. In addition, $\mathbf B_t^0$ can be solved from the ODEs \eqref{odeB}
and \eqref{eqn:fB} if it is needed.

In the regime of slow dislocation reaction $M_n/M_B\gg l^2/B^2$, the grain boundary dynamic equations of $\mathbf v$ and $\dot{\theta}$ are still
Eqs.~\eqref{eqn:v-slow} and \eqref{eqn:ll6}, with $\mathbf v^*$  given by Eq.~\eqref{eqn:v-star-stress},
\begin{flalign}
\mathbf B^0_t=&\frac{M_B}{2\sin\frac{\theta}{2}}\left[(\gamma\mathbf n-\gamma' \mathbf T)
+\int_0^s\big(\pmb \sigma\cdot\mathbf B+\Psi\mathbf T\big)dw \right]+\mathbf C_B,
\end{flalign}
where  the constant $\mathbf C_B$ is a constant such that Eq.~\eqref{eqn:bzeroaverage} holds, and recall that $S$ is calculated from  $\mathbf B^0_t$ by Eq.~\eqref{eqn:Cs}. In this case, the stress $\pmb\sigma$ and the synthetic force $\Psi$ contribute to all the three variables $\dot{\theta}$, $\mathbf B^0_t$, and $\mathbf v$.

In the regime of fast dislocation reaction $M_n/M_B\ll l^2/B^2$, we have
\begin{flalign}
\mathbf v=&M_n\big[(\gamma+\gamma'')\kappa \mathbf n+(\pmb \sigma\cdot\mathbf B)\times\hat{z} -\Psi\mathbf n \big], \label{eqn:fastapp1}\\
\dot{\theta} =&-\frac{M_\theta}{l}\int_\Gamma\frac{\partial \gamma}{\partial \theta} ds.\label{eqn:fastapp2}
\end{flalign}
Here we have the contributions of the stress $\pmb\sigma$ and the synthetic force $\Psi$ in the grain boundary velocity, in addition to the velocity formulations of the
 available grain boundary dynamic models with grain boundary sliding \cite{Li1962,Shewmon1966,Kobayashi2000,Upmanyu2006,Selim2016,epshteyn2019motion,epshteyn2019large}.

Note that the grain boundary dynamic equations \eqref{eqn:fastapp1} and \eqref{eqn:fastapp2} in the fast dislocation reaction regime also hold for an infinite grain boundary if the average in Eq.~\eqref{eqn:fastapp2} is finite.
For example, for an infinite planar low angle tilt boundary parallel to the $y$ axis with normal direction $\mathbf n$ in the $-x$ direction.
Its dislocation density is $\mathbf B=-2\sin\frac{\theta}{2}\mathbf n=\sin\frac{\theta}{2}\hat{\mathbf x}$, where $\hat{\mathbf x}$ is the unit vector in $+x$ direction. This dislocation density is due to an edge dislocation array with Burgers vector $\mathbf b=b\hat{\mathbf x}$ and inter-dislocation distance $D=b/(2\sin\frac{\theta}{2})$.
When the misorientation angle does not change, i.e., $\dot{\theta}=0$, under a constant  shear stress $\tau=\sigma_{12}$, the velocity of the grain boundary using our formulation is $\mathbf v=\mathbf v^*=M_n\tau b\hat{\mathbf x}$. This agrees with the classical model for the motion of a tilt boundary~\cite{HL,Sutton1995}. Note that by Eq.~\eqref{eqn:parallel},
the  tangential relative velocity in this case is $v_\parallel= 2\tan\frac{\theta}{2} v_n$, which recovers the result in the Cahn-Taylor theory~\cite{Cahn20044887} with the shear coupling factor $\beta=2\tan\frac{\theta}{2}$.

\section{High angle grain boundaries with disconnection structure}\label{sec:high-angle}

\begin{figure}[htbp]
\centering
\subfigure[]    {\includegraphics[width=.7\linewidth]{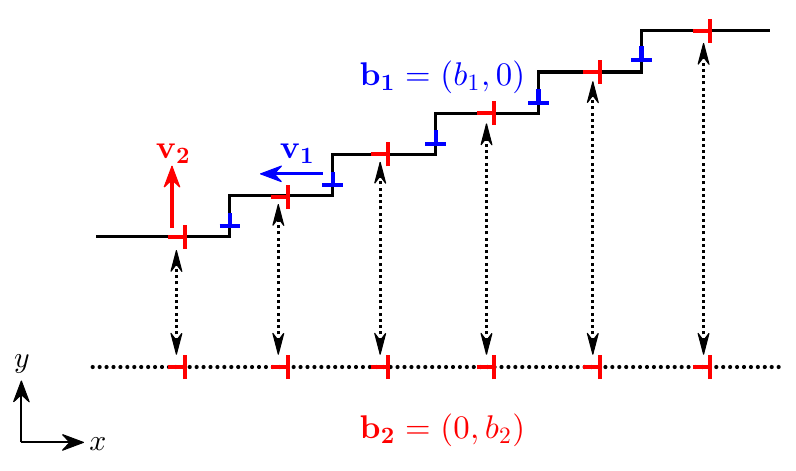}}
\subfigure[]    {\includegraphics[width=.7\linewidth]{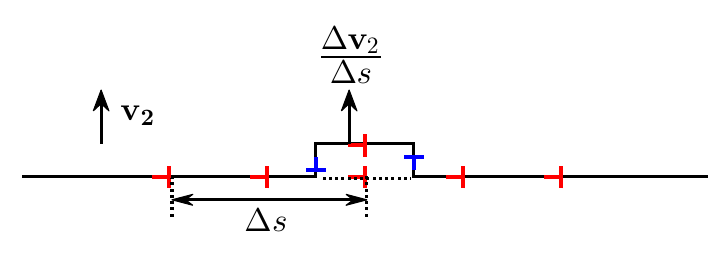}}
    \caption{(a) A high angle grain boundary with structure of disconnections and dislocations. Each disconnection has both a Burgers vector $\mathbf b_1$ and a step height $H_1$, and a velocity $\mathbf v_1$. The Burgers vector $\mathbf b_1$ and the velocity $\mathbf v_1$ are in the direction parallel to the reference plane (i.e. the $x$ direction). The dislocations of this high angle grain boundary are those associated with the reference plane, and have the Burgers vector $\mathbf b_2$, and their velocity $\mathbf v_2$  in the direction normal to the reference plane (i.e. the $y$ direction). It is assumed that the dislocation velocity $\mathbf v_2$ is associated only with nucleation of disconnections as shown in (b). (b) Nucleation of disconnections is only associated with variation of the velocity  of the grain boundary (and the dislocations) normal to the reference plane ($\frac{dv_2}{ds}$), and vice versa.}
    \label{fig:demo2}
\end{figure}

In the variation formulation presented in the previous sections, the microstructure of a grain boundary is described by $\mathbf B$, which is the density of the constituent dislocations for a low angle grain boundary and the density of closure failure of the Burgers circuit for a high angle grain boundary. These are based on the classical grain boundary theory~\cite{HL,Sutton1995}.
In the recently developed disconnection theories of high angle grain boundaries \cite{Ashby1972498,King1980335,Hirth19964749,Rajabzadeh20131299,Thomas2017,Zhang2017119,Han2018386,Wei2019133,Zhang-npj2021,Han2021,Han2024}, a high angle grain boundary has a structure of disconnections relative to a reference planar grain boundary; see Fig.~\ref{fig:demo2}(a).
A disconnection is a line defect that has both a Burgers vector $\mathbf b$ and a step height $H$.
In this section, we generalize the variational model to high angle grain boundaries with these disconnection structure.

Assuming that the reference planar grain boundary has normal direction $(0,1)$, and misorientation angle $\theta$.   The reference plane always satisfies the Frank-Bilby equations, i.e., the dislocation density of the reference plane  $\mathbf B^{\rm ref}$ satisfies
\begin{eqnarray}\label{eqn:high1}
 \mathbf B^{\rm ref}=(B_1^{\rm ref},B_2^{\rm ref})=-2\sin\frac{\theta}{2}(0,1).
\end{eqnarray}
Note that when the reference planar grain boundary is low angle, $\mathbf B^{\rm ref}$ is the dislocation density, otherwise $\mathbf B^{\rm ref}$ is the density of closure failure as discussed in Sec.~\ref{sec:constraints}.

Consider a high angle grain boundary with respect to this reference plane. The grain boundary can be described by $y=h(x)$, where the slope $h_x$ is small. The disconnections of the high angle grain boundary  have Burgers vector $\mathbf b=(b_1,0)$ and step height $H_1$. The disconnection number density along the $x$ direction is $-\frac{1}{H_1}h_x$, and accordingly, the disconnection density along the grain boundary is $B_1\hat{\mathbf x}$, with
\begin{eqnarray}\label{eqn:high2}
B_1=-\frac{b_1}{H_1}h'(x)\frac{1}{\sqrt{1+h_x^2}}=-\beta_1 n_1,
\end{eqnarray}
where $\beta_1=\frac{b_1}{H_1}$, and $\mathbf n=(n_1,n_2)=\frac{1}{\sqrt{1+h_x^2}}(h_x,-1)$ is the normal direction of the grain boundary.

This high angle grain boundary should inherit the dislocation density on the reference plane. Considering both Eqs.~\eqref{eqn:high1} and \eqref{eqn:high2}, on this grain boundary, we have
\begin{eqnarray}\label{eqn:high3}
\mathbf B=(B_1,B_2)=(\beta_1 n_1,-2\sin\frac{\theta}{2} n_2).
\end{eqnarray}
See the schematic illustration in Fig.~\ref{fig:demo2}(a) for the disconnection structure of the high angle grain boundary.
We call Eq.~\eqref{eqn:high3} \underline{\bf unified Frank-Bilby equations}.


The grain boundary velocity is $\mathbf v=(v_1,v_2)$, where $v_1$ is from the motion of disconnections in the $x$ direction, and $v_2$ is from nucleation of disconnections. See the schematic illustration in Fig.~\ref{fig:demo2}(a).

Here the  high angle grain boundary model in Eq.~\eqref{eqn:high3} and the dynamics described above are based on the underlying microstructure and dynamics of both disconnections and dislocations.
The structure and dynamics of disconnections are consistent with the available continuum models \cite{Zhang2017119,Wei2019133,Zhang-npj2021,Han2021,Han2024} and  atomistic simulations \cite{Thomas2017}. Note that $\beta_1$ could be either positive or negative. 
The dislocation structure of the reference grain boundary satisfies
the Frank-Bilby equations, which agrees with the available dislocation based continuum models for grain boundary dynamics \cite{Zhang2018157,zhang2019new,Qin2021}; however, unlike in those models, here these dislocations in the high angle grain boundary
in general do not move and the dislocation motion is related only to the nucleation of disconnections,
which are consistent with the available disconnection based models and simulations \cite{Thomas2017,Zhang2017119,Han2018386,Wei2019133,Zhang-npj2021,Han2024}.
Note that in \cite{Han2021,Han2024}, high angle grain boundaries with multiple reference planes are considered for the dynamics and grain rotation, and they did not consider the equilibrium dislocation structure on the reference planes. Here we carefully examine the high angle grain boundary with a single reference plane, and incorporate the dislocation structure of the reference plane which is associated with the misorientation angle $\theta$, whose evolution leads to grain rotation. 
Our model here can be considered as a basis for the multi-reference grain boundary dynamics models in \cite{Han2021,Han2024} and other grain boundary dynamics models coarse-grained at larger length/time scales from the microscopic models.

During the motion of the high angle grain boundary, the change of grain boundary normal direction $\dot{\mathbf n}$ is given by Eq.~\eqref{eqn:ds}, and
the dynamics of disconnection/disloction densities $\dot{\mathbf B}=(\dot{B}_1,\dot{B}_2)$  are given by
\begin{flalign}
\dot{B}_1&=-\left(\frac{d\mathbf v}{ds}\cdot\mathbf T\right)B_1+{B_1}_t^0,\label{eqn:db1}\\
\dot{B}_2&=-\left(\frac{d\mathbf v}{ds}\cdot\mathbf T\right)B_2-\frac{dv_1}{ds}2\sin\frac{\theta}{2}+{B_2}_t^0,\label{eqn:db2}
\end{flalign}
where $\mathbf B^0_t=({B_1}_t^0,{B_2}_t^0)$ with ${B_1}_t^0$ and ${B_2}_t^0$ being
 the changes of disconnection and dislocation contents due to their reactions (nucleation or annihilation) per unit length  on the grain boundary per unit time.
Note the Lagrange coordinate frame in terms of the disconnections $B_1$ is used in these equations.
Compared with the corresponding dynamics of dislocation densities  in Eq.~\eqref{eqn:B0t0},
the  new, second term in Eq.~\eqref{eqn:db2} is to include additional $B_2$ dislocations per unit length in the moving frame of the disconnections $B_1$ due to their motion, which is
$\frac{dv_1}{ds}B_2/n_2=-\frac{dv_1}{ds}2\sin\frac{\theta}{2}$,
 because $B_2$ dislocations do not move with the disconnections $B_1$.

The dynamic Frank-Bilby equations for this high angle grain boundary  are $\mathbf h_d=( {h_d}_1, {h_d}_2, {h_d}_3)=\mathbf 0$, where
\begin{flalign}
{h_d}_1=&\frac{dv_2}{ds}\beta_1+{B_1}_t^0=0,\label{eqn:dh0}\\
{h_d}_2=&\dot{\theta}\cos\frac{\theta}{2}n_2+{B_2}_t^0=0,\label{eqn:dh2}\\
{h_d}_3=&\int_\Gamma \left(-\dot{\theta}y(s)n_2-k\mathbf v\cdot \mathbf n\right)ds
=\dot{\theta}A-\int_\Gamma k \mathbf v\cdot \mathbf n ds=0,
\label{eqn:dh1}
\end{flalign}
 where $k=\beta_1+2\sin\frac{\theta}{2}$, and
 $A=-\int_\Gamma y(s)n_2ds$ is  the  area  enclosed  by  the  grain  boundary $\Gamma$.

Equations  \eqref{eqn:dh0} and \eqref{eqn:dh2}  are obtained by taking time derivative in Eq.~\eqref{eqn:high3}, and further use Eqs.~\eqref{eqn:high3}-\eqref{eqn:db2}, and \eqref{eqn:ds}.
The meaning of Eq.~\eqref{eqn:dh0} is that nucleation of disconnections is only associated with the nonuniform velocity of the dislocations in the direction normal to the reference plane, and vice versa; see Fig.~\ref{fig:demo2}(b), which describes the nature of a disconnection with both characters of a Burgers vector and a step. The meaning of Eq.~\eqref{eqn:dh2} is the change of misorientation angle (i.e., grain rotation) is due to the change of dislocation density, which is also the equilibrium dislocation density of the reference plane.

Eq.~\eqref{eqn:dh1} is based on the plastic flow generated by the motion of disconnections due to their character of Burgers vector. In fact, the grain rotation by $\dot{\theta}$ needs a relative tangential velocity $-\dot{\theta}y(s)$ in the $x$ direction, if it is generated by the plastic flow $v_1B_1$, we have
$-\dot{\theta}y(s)=v_1B_1/(-n_2)$. Taking derivative with respect to the arclength $s$, and incorporating the vertical velocity of the dislocation and the reaction rate ${B_1}_t^0$ of Burgers vectors in the $\mathbf x$ direction (following Eq.~\eqref{eqn:FBderivative18}), which are the Burgers vectors of the disconnections, we have
\begin{flalign}
\dot{\theta}n_1-\beta_1\frac{d}{ds}\left(v_1\frac{n_1}{n_2}\right)-2\sin\frac{\theta}{2}\frac{dv_2^{\rm total}}{ds}+{B_1}_t^0=0.
\end{flalign}
Here $v_2^{\rm total}$ is the actual vertical velocity of the dislocations.
\begin{flalign}
v_2^{\rm total}=v_1\frac{n_1}{n_2}+v_2. \label{eqn:v2total}
\end{flalign}
Using Eqs.~\eqref{eqn:dh0} and \eqref{eqn:v2total}, we have
\begin{flalign}
\dot{\theta}n_1-k\frac{dv_2^{\rm total}}{ds}=0.
\end{flalign}
Integrating it with respect to the arclength $s$ and using Eq.~\eqref{eqn:v2total}, the relative tangential velocity is
\begin{flalign}
-\dot{\theta}y(s)n_2=kv_2^{\rm total}n_2=k\mathbf v\cdot \mathbf n. \label{eqn:dh1-exact}
\end{flalign}
This equation associated the motion of disconnections and grain rotation should hold in some average sense since instant equilibrium is not required. This leads to Eq.~\eqref{eqn:dh1}.

Eq.~\eqref{eqn:dh1} (or the exact form \eqref{eqn:dh1-exact}) means that the normal velocity generate a relative tangential velocity with coupling factor $k=\beta_1+2\sin\frac{\theta}{2}$, recalling that $\beta_1=b_1/H_1$. Note that the contribution $b_1$ in the coupling factor $k$ comes from the horizontal motion of the disconnections, while the contribution of $2\sin\frac{\theta}{2}$ in the coupling factor $k$  is due to the vertical motion of the dislocations associated with the reference state. Note that in the recent disconnection based grain boundary rotation model \cite{Han2024}, the coupling factor is $\beta_1$, which only considers the relative tangential motion (grain rotation) due to the motion of disconnections. For a special case that the initial disconnection density is in equilibrium, i.e., there is no long-range stress field associated with disconnections, the Frank-Bilby equations \eqref {eqn:B0} hold for both disconnection density $B_1$ and dislocation density $B_2$, and we have $\beta_1=-2 \sin\frac{\theta}{2}$; in this case, $k=0$, which means that the motion of such a grain boundary does not lead to relative tangential motion or grain rotation.

As in the formulation presented in the previous sections, this unified variational formulation for the  dynamics of a high angle grain boundary consists of evolution of the grain boundary and evolution of the disconnection and dislocation contents $\mathbf B=(B_1,B_2)$ on the grain boundary. The formulation is based on the Onsager principle~\cite{Onsager,Doi2015} and constrained minimization subject to the dynamic Frank-Bilby equations
in Eqs.~\eqref{eqn:dh0}--\eqref{eqn:dh1}.  The independent variables are the grain boundary velocity $\mathbf v=(v_1,v_2)$ (i.e., the disconnection velocity $v_1$ and dislocation velocity $v_2$), and  the reaction rates of disconnection and dislocation contents  $\mathbf B_t^0=({B_1}_t^0, {B_2}_t^0)$, together with the dynamics of   disconnection and dislocation contents  $\mathbf B$ in \eqref{eqn:db1} and \eqref{eqn:db2} and the unified Frank-Bilby equations \eqref{eqn:high3}.

The constrained minimization problem is
\begin{eqnarray}
{\rm minimize \ } &&
 Q+\dot{E} \vspace{1ex}\label{eqn:onsagerd}\\
{\rm subject \ to \ }&&  \mathbf h_d=\mathbf 0, \label{eqn:onsager-hd}
\end{eqnarray}
where the constraints $\mathbf h_d=\mathbf 0$ are given in Eqs.~\eqref{eqn:dh0}--\eqref{eqn:dh1}.

In the minimization problem,
$Q$ is the dissipation function:
\begin{flalign}\label{eqn:dissipationd}
Q=\int_\Gamma \left(\frac{1}{2{M_n}_1}{v_1}^2+\frac{1}{2{M_n}_2}{v_2}^2
+\frac{1}{2{M_B}_1}({B^0_1}_t)^2+\frac{1}{2{M_B}_2}({B^0_2}_t)^2
+\frac{1}{2M_\theta}\dot{\theta}^2
\right) ds,
\end{flalign}
where ${M_n}_1$ and ${M_n}_2$ are mobilities of the grain boundary velocity component $v_1$ and $v_2$, respectively, ${M_B}_1$ and ${M_B}_2$ are mobilities of
the reaction rates of disconnection and dislocation contents $B_1$ and $B_2$, respectively, ${M_n}_1$ and ${M_n}_2$,  ${M_B}_1$ and ${M_B}_2$ may be different following the microstructure and dynamics of the high angle grain boundary described above,
 and $M_\theta$ is the mobility for grain rotation $\dot{\theta}$.
The term
$\dot{E}$ is the dissipation rate of the total energy $E$, 
and for this high angle grain boundary,
the rate of energy dissipation is
\begin{flalign}\label{eqn:energy-dissipationd}
 \dot{E}=&\int_\Gamma\left[\gamma(\mathbf B)\left(\frac{d\mathbf v}{ds}\cdot\mathbf T\right)+\frac{\partial \gamma}{\partial \mathbf B}
 \cdot\left(-\left(\frac{d\mathbf v}{ds}\cdot\mathbf T\right)\mathbf B+\left(0,-\frac{dv_1}{ds}2\sin\frac{\theta}{2}\right)  +\mathbf B^0_t\right) \right.\nonumber\\
 & +\sigma_{12}\left(-B_1v_1+B_2v_1\frac{n_1}{n_2} +B_2v_2\right)+ \Psi (\mathbf v\cdot \mathbf n)\bigg]ds.
\end{flalign}
Here we have used the expressions of $\dot{ds}$ and $\dot{\mathbf B}$ in Eqs.~\eqref{eqn:dss}, \eqref{eqn:db1}, and \eqref{eqn:db2}.
Compared with the energy dissipation formula  in Eq.~\eqref{eqn:energy-dissipation1} for low angle grain boundary with dislocation structure, here there is an additional term $\left(0,-\frac{dv_1}{ds}2\sin\frac{\theta}{2}\right)$ to account for the disconnection/disloction structure of the high angle grain boundary, following Eq.~\eqref{eqn:db2}. The contributions of stress $\pmb \sigma$ in $\dot{E}$ due to motions of disconnections and dislocations, i.e. $-\sigma_{12}B_1v_1$ and $\sigma_{12}B_2\left(v_1\frac{n_1}{n_2}+v_2\right)=\sigma_{12}B_2v_2^{\rm total}$, respectively, are calculated from the Peach-Koehler force on dislocations $((\pmb \sigma\cdot\mathbf B)\times \hat{\mathbf z})$ \cite{HL}. Note that the actual velocity of a dislocation in the vertical direction is $v_2^{\rm total}$.

Now we solve the constrained minimization problem of Eqs.~\eqref{eqn:onsagerd} and \eqref{eqn:onsager-hd}.
The Lagrangian function of the minimization problem with the constraint $\mathbf h_d=\mathbf 0$ in \eqref{eqn:onsager-hd} is
\begin{flalign}\label{eqn:Lagranged}
L=&Q+\dot{E}+\int_\Gamma \pmb\lambda \cdot \mathbf h_d ds,
\end{flalign}
where $\pmb \lambda=(\lambda_1,\lambda_2,\lambda_3)$.
Taking variations of this Lagrangian function, we have
\begin{flalign}
\frac{\delta L}{\delta v_1}=&\frac{1}{{M_n}_1}v_1-\frac{1}{{M_n}_1}v^*_1-\lambda_3 kn_1=0,\label{eqn:ld1}\\
\frac{\delta L}{\delta v_2}=&\frac{1}{{M_n}_2}v_2-\frac{1}{{M_n}_2}v^*_2-\beta_1\frac{d\lambda_1}{ds}-\lambda_3kn_2=0,\label{eqn:ld2}\\
\frac{\delta L}{\delta  {B^0_1}_t}=& \frac{1}{{M_B}_1}{B^0_1}_t-\frac{1}{{M_B}_1}{B_1}_t^{0*}+\lambda_1=0,\label{eqn:ld3}\\
\frac{\delta L}{\delta  {B^0_2}_t}=& \frac{1}{{M_B}_2}{B^0_2}_t-\frac{1}{{M_B}_2}{B_2}_t^{0*}+\lambda_2= 0,\label{eqn:ld4}\\
\frac{\delta L}{\delta \dot{\theta}}=&\frac{l}{M_\theta} \dot{\theta}
+\cos\dfrac{\theta}{2}\int_\Gamma\lambda_2n_2 ds+\lambda_3A=0.\label{eqn:ld5}
\end{flalign}
These equations should be solved together with the constraint $\mathbf h_d=\mathbf 0$ in \eqref{eqn:onsager-hd}, i.e.,
Eqs.~\eqref{eqn:dh0}--\eqref{eqn:dh1}, and  the unified Frank-Bilby equations \eqref{eqn:high3}.
Here for the high angle grain boundary, $\mathbf v^*=(v_1^*,v_2^*)$ and $\mathbf B_t^{0*}=( {B_1}_t^{0*},{B_2}_t^{0*})$ are given by 
\begin{flalign}
\mathbf v^*=&-\mathbf M_n\frac{\delta \dot{E}}{\delta \mathbf v} \label{eqn:vd*}\\
=&\mathbf M_n \frac{d}{ds}\left[\left(\gamma(\mathbf B)-\frac{\partial \gamma}{\partial \mathbf B}\cdot \mathbf B\right)\mathbf T+\left(-2\sin\frac{\theta}{2}\frac{\partial \gamma}{\partial  B_2},0\right)\right.\\
&\left.+\sigma_{12}\left(B_1-B_2\frac{n_1}{n_2},-B_2\right) -\Psi\mathbf n\right],\nonumber \\
\mathbf {B^0_t}^*=&-\mathbf M_B \frac{\delta \dot{E}}{\delta \mathbf B^0_t}=-\mathbf M_B\dfrac{\partial \gamma}{\partial \mathbf B},\label{eqn:Bd*}
\end{flalign}
with mobility matrices $\mathbf M_n$ and $\mathbf M_B$ being diagonal matrices $\mathbf M_n={\rm diag}({M_n}_1,{M_n}_2)$ and $\mathbf M_B={\rm diag}({M_B}_1,{M_B}_2)$.
Note again that compared with the unconstrained velocity $v^*$ for low angle grain boundary in Eq.~\eqref{eqn:v-star-stress},
the additional term $\left(0,-\frac{dv_1}{ds}2\sin\frac{\theta}{2}\right)$ in Eq.~\eqref{eqn:vd*} is to account for the disconnection/disloction structure of the high angle grain boundary, following Eq.~\eqref{eqn:db2}.

We solve these equations under the assumption
${M_n}_2/ {M_B}_1\ll L^2$, where $L$ is the length scale of the continuum model, which means that
the nucleation of disconnections is much faster than the motion of dislocations in the vertical direction (i.e., the direction normal to the reference plane),
and the vertical velocity $v_2$ is mainly from nucleation of disconnections instead of motion of dislocations in the vertical direction; cf. Eq.~\eqref{eqn:dh1}. We also assume that $M_\theta\ll {M_B}_2$, i.e. fast dislocation reaction.

Under the above conditions, from Eqs.~\eqref{eqn:ld1}--\eqref{eqn:ld5}, \eqref{eqn:dh0}--\eqref{eqn:dh1}, \eqref{eqn:high3},
and \eqref{eqn:vd*}, \eqref{eqn:Bd*},  we have
\begin{flalign}
v_1
= &v_1^0+\lambda_3k{M_n}_1n_1\vspace{1ex}\label{eqn:v_1disconnection}\\
v_2=&v_2^0+\lambda_3k{M_n}_2n_2,\label{eqn:v_2disconnection}
\end{flalign}
with $\lambda_3$ solved from \eqref{eqn:dh1} and $\dot{\theta}$ from (6.22):
\begin{flalign}
\lambda_3=&\frac{1}{k\big({M_n}_1\int_\Gamma n_1^2ds+{M_n}_2\int_\Gamma n_2^2ds\big)}\left(
-\int_\Gamma \mathbf v^0\cdot \mathbf n ds+\frac{A}{k}\dot{\theta}\right), \vspace{1ex}\label{eqn:lambda1_fina3}\\
=&-\frac{\frac{l}{A^2}}{M_\theta+\frac{k^2l}{A^2}\big({M_n}_1\int_\Gamma n_1^2ds+{M_n}_2\int_\Gamma n_2^2ds\big)}\left(\int_\Gamma k\mathbf v^0\cdot \mathbf n ds+M_\theta A\frac{\partial \gamma}{\partial \theta}\right)\vspace{1ex}\nonumber\\
\dot{\theta}=&\frac{1}{A}\int_\Gamma k \mathbf v\cdot \mathbf n ds\label{eqn:theta-high1}\\
=&\frac{M_\theta}{M_\theta+\frac{k^2l}{A^2}\big({M_n}_1\int_\Gamma n_1^2ds+{M_n}_2\int_\Gamma n_2^2ds\big)}\vspace{1ex}
\nonumber\\
& \cdot\left[\frac{1}{A}\int_\Gamma k\mathbf v^0\cdot\mathbf n ds-\frac{k^2l}{A^2}\left({M_n}_1\int_\Gamma n_1^2ds+{M_n}_2\int_\Gamma n_2^2ds\right)\frac{\partial \gamma}{\partial \theta}\right], \nonumber
\end{flalign}
where the $\mathbf v^0=(v_1^0,v_2^0)$ is the grain boundary velocity with fast dislocation and disconnection reactions presented in the previous section:
\begin{flalign}
v_1^0=&v_1^*={M_n}_1\big[(\gamma+\gamma'')\kappa n_1+\sigma_{12}(B_1-B_2{\textstyle \frac{n_1}{n_2}}) -\Psi n_1\big],\label{eqn:v1000}\\
v_2^0=&v_2^*+{\textstyle \frac{{M_n}_2\beta_1}{{M_B}_1}\frac{d {B_1}_t^{0*}}{ds}}={M_n}_2\big[(\gamma+\gamma'')\kappa n_2 -\sigma_{12}B_2 -\Psi n_2\big]. \label{eqn:v2000}
\end{flalign}
Here  in the calculation of $\dot{\theta}$, we have used $\frac{\partial \gamma}{\partial \theta}=-\cos\frac{\theta}{2}\frac{\partial \gamma}{\partial B_2}n_2$ based on the unified Frank-Bilby equations \eqref{eqn:high3}.

In numerical simulations, it is more convenient to use $x$ as the parameter of the grain boundary, and the evolution of the grain boundary can be written as
\begin{flalign}
h_t=v_2^{\rm total}=v_1\frac{n_1}{n_2}+v_2, \label{eqn:v2total-ht}
\end{flalign}
with $v_1$ and $v_2$ given by  Eqs.~\eqref{eqn:v_1disconnection} and \eqref{eqn:v_2disconnection}, and grain rotation $\dot{\theta}$ by Eq.~\eqref{eqn:theta-high1}.

Consider the grain rotation formula in Eq.~\eqref{eqn:theta-high1}, which is caused by the relative tangential motion of the grain boundary that is proportional to the normal velocity of the grain boundary. As discussed above, the coupling factor $k=\beta_1+2\sin\frac{\theta}{2}$ includes both the coupling effects from the horizontal motion of the disconnections ($\beta_1$) and the vertical motion of the dislocations associated with the reference state ($2\sin\frac{\theta}{2}$). The physical origins of grain rotation can be seen from the last equation in \eqref{eqn:theta-high1}.
The first term in it that depends on $k\mathbf v^0\cdot\mathbf n=k(v^0_1n_1+v^0_2n_2)$ comes from the relative tangential motion of the grain boundary without grain rotation constraint, which is proportional to the normal velocity of the grain boundary due to the motions of disconnections in the $x$ direction ($v_1n_1$)  and the nucleation/annihilation of disconnections which causes motion of the grain boundary in the $y$ direction due to the step character of the disconnections ($v_2^0n_2$, cf. \eqref{eqn:dh0}). The the second term in the last equation in \eqref{eqn:theta-high1} that depends on $\frac{\partial \gamma}{\partial \theta}$  is from the reaction of the dislocations that determine the misorientation angle of the grain boundary (cf. \eqref{eqn:dh2}).

Note that the above equations for the grain boundary dynamics and grain rotation  
also hold for a grain with multiple reference planes, where the disconnections and dislocations as well as their velocities and reaction rates of the high angle grain boundary in different inclination angles may have different reference planes.

This grain boundary velocity $\mathbf v$ 
agrees with the disconnection based high angle grain boundary dynamics model presented in \cite{Zhang2017119} when grain rotation is not considered, i.e. $\lambda_3=0$.
In fact, using the Eulerian coordinates $(x,h(x))$ of the grain boundary, the grain boundary velocity is given by $v_2^{\rm total}$ in $y$ direction in Eq.~\eqref{eqn:v2total}.
 When ${M_n}_2=0$, i.e., only the disconnections are allowed to move in the horizontal direction, using Eq.~\eqref{eqn:v2total-ht}, we have
\begin{flalign}
h_t
 \approx &-\frac{{M_n}_1|n_1|}{H_1} \big[-(\gamma+\gamma'')h_{xx} H_1 +\sigma_{12}b_1{\textstyle \frac{k}{\beta_1}} -\Psi H_1\big]|h_x|.
\end{flalign}
Here we have used $B_1=\beta_1 n_1$, $\beta_1=\frac{b_1}{H_1}$, $k=\beta_1+2\sin\frac{\theta}{2}$, $\frac{n_1}{|n_2|}\approx h_x$, $\kappa\approx-h_{xx}$ for small $h_x$. This normal velocity without nucleation of disconnections agrees with that in \cite{Zhang2017119}  if the disconnection mobility there is set as $M_d=\frac{{M_n}_1|n_1|}{H_1}$, and $k\approx \beta_1$ (i.e., the initial deviation in  misorientation angle from the reference state is negligible). When ${M_n}_2\neq 0$, following  Eq.~\eqref{eqn:v2total-ht}, there is an additional contribution  of $v_2^0$ given by Eq.~\eqref{eqn:v2000} in $h_t$,
which is consistent with the small constant density of equilibrium disconnection pairs in the velocity in \cite{Zhang2017119} except that driving force due to vertical motion of the associated dislocations under the stress $\sigma_{12}$ is more accurately accounted in our formulation presented here for the vertical velocity due to nucleation of disconnections.

When grain rotation is not allowed, i.e., $\dot{\theta}=0$ or $M_\theta=0$, 
and further, the disconnections are only allowed to move in the horizontal direction, i.e. ${M_n}_2=0$,
using the Eulerian coordinates $(x,h(x))$ of the grain boundary as above, we have
\begin{flalign}
h_t \approx &-\frac{{M_n}_1|n_1|}{H_1} \big[-(\gamma+\gamma'')h_{xx} H_1 +\sigma_{12}b_1 -\Psi H_1+\lambda_3k\big]|h_x|,
\end{flalign}
where the constant contribution  $\lambda_3k=-\frac{1}{{M_n}_1\int_\Gamma n_1^2ds}\int_\Gamma \mathbf v^0\cdot \mathbf n ds$ is from the Lagrange multiplier.
Here we also assume that $k\approx \beta_1$, i.e., the initial deviation in  misorientation angle from the reference state is negligible.
This equation is similar to that obtained in \cite{Zhang-npj2021} for the case of no tangential plastic deformation and under the condition  ${M_n}_2=0$  considered in \cite{Zhang-npj2021}, in which the stress field is adjusted through balance of different disconnection modes due to the constraint.

Compared with another recently proposed disconnection based grain rotation model \cite{Han2024}, there are mainly the following three differences between our model and that in \cite{Han2024}. Firstly, misorientation change of the reference planar grain boundary is included in our model by considering the equilibrium dislocation densities on the reference plane determined by Frank-Bilby equations,
which provides a well-defined misorientation angle and misorientation dependent energy density for the grain boundary; while in the model in  \cite{Han2024}, disconnections with multiple reference planes are considered.
Secondly, in our grain rotation formula \eqref{eqn:theta-high1}, grain rotation angle is calculated directly based on the quantities of the grain boundary, including contributions from velocity and dislocation/disconnection reaction (misorientation dependent grain boundary energy density); whereas in \cite{Han2024}, rotation angle is calculated over the entire domain and the value at the grain center is chosen as the grain rotation angle. For the rectangular grain examined in \cite{Han2024}, our grain rotation formula \eqref{eqn:theta-high1} gives similar result as theirs.
Thirdly, our model considers explicitly the change of disconnection and dislocation densities due to change of arclength and that due to nucleations and reactions; whereas in \cite{Han2024}, a single mobility is used to account for the changes from both effects.
Note that in our model, we consider disconnections with a single model in order to capture the natures of microstructure dependent grain boundary motion in a mathematically tractable theory. Multi modes of disconnections \cite{Wei2019133,Han2021,Han2024} can also be incorporated in our continuum model.

The stress component $\sigma_{12}$ in our formulation includes both the self stress from the disconnections and the applied stress,
 i.e. $\sigma_{12}=\sigma_{12}^{\rm self}+\sigma_{12}^{\rm appl}$. Note that unlike the cancellation of the self stress field of dislocations  due to the equilibrium (Frank-Bilby equations) in the formulation for low angle grain boundaries in the previous sections, here the self stress field of disconnections are present because equilibrium of disconnection distributions are not required. Using the  coordinates $(x,h(x))$ of the grain boundary, the self stress  is \cite{Zhang2017119}
\begin{equation}
\sigma_{12}^{\rm self}(x,t)=\int\frac{\mu}{2\pi(1-\nu)}\frac{\beta_1h_x(x_1,t)}{x-x_1}dx_1.
\end{equation}
In a general case, on a point $(x,y)$ on the grain boundary $\Gamma$, the self stress is
$\sigma_{12}^{\rm self}(x,y,t)=\int_\Gamma \frac{\mu}{2\pi(1-\nu)}\frac{(x-x_1)[(x-x_1)^2-(y-y_1)^2]}{[(x-x_1)^2+(y-y_1)^2]^2}B_1(x_1,y_1,t)ds$, where $(x_1,y_1)$ is the point varying on the grain boundary $\Gamma$. There is no self stress associated with the dislocation distribution $B_2$ because of the equilibrium assumption.

\section{Conclusions and discussion}\label{sec:conclustions}

In this paper, we have proposed a unified variational framework to account for all the underlying line defect mechanisms (dislocations and disconnections) for the dynamics of both low and high angle grain boundaries.   The variational formulation is based on the developed  constraints of the dynamic Frank-Bilby equations that govern the microscopic structures.
 The variational framework incorporates all kinds of motions, including the coupling motion as well as motion by mean curvature and the sliding motion,
  and the driving forces due to grain boundary energy density, the  stress, and chemical potential jump across the grain boundary.
   The variational framework is able to recover the available models for these different motions under different conditions. 
Numerical simulations are performed to validate the continuum formulations resulting from the proposed variational framework.

The unified variational framework will be more efficient to describe the collective behaviors of grain boundary networks at larger length scales. It will also provide a basis for rigorous analysis of these partial differential equation models and for the development of efficient numerical methods.

The developed variational framework also applies to dynamics of
semicoherent hetero-interfaces of bicrystals with dislocation strucure, for which the Frank-Bilby equations are \cite{HL,Sutton1995,Frank1950,Bilby1955}:
\begin{eqnarray}\label{eqn:FBhetero}
\mathbf B_p=2\sin\frac{\theta}{2}(\mathbf p\times \mathbf a)+\Sigma\mathbf p,
\end{eqnarray}
where matrix $\Sigma=S_\beta^{-1}-S_\alpha^{-1}$ with $S_\alpha$ and $S_\beta$ respectively being the distortion transformation matrices  that map the
lattice vectors from the natural unstrained crystal lattices $\alpha$ and $\beta$ to the reference
lattice.
In the two-dimensional setting,  the Frank-Bilby equations in Eq.~\eqref{eqn:FBhetero} are
\begin{eqnarray}
 \mathbf B=-2\sin\frac{\theta}{2}\mathbf n+\Sigma\mathbf T.
\end{eqnarray}
From Eq.~\eqref{eqn:ds},   we have $\dot{\mathbf T}=\left(\frac{d\mathbf v}{ds}\cdot\mathbf n\right)\mathbf n$. Thus, taking time derivative in the Frank-Bilby equations \eqref{eqn:FBhetero}, similarly to Eq.~\eqref{eqn:FBderivative18} , we have the dynamic Frank-Bilby equations in this case:
\begin{eqnarray}
\mathbf h_{\rm hetero}=\dot{\theta}\cos\frac{\theta}{2}\mathbf n- \left(\frac{d\mathbf v}{ds}\cdot\mathbf n\right)\Sigma\mathbf n-2\sin\frac{\theta}{2}\frac{d}{ds}
\left(\mathbf v\times \hat{\mathbf z}\right)+\mathbf B^0_t=\mathbf 0.
\end{eqnarray}
For a high angle hetero-interface with disconnection and dislocation structures under the same conditions as the high angle grain boundary discussed in Sec.~\ref{sec:high-angle}, there is an additional term
depending on $\Sigma$ in Eqs.~\eqref{eqn:dh2} and \eqref{eqn:dh1} for the dynamic Frank-Bilby equations. Note that the exact variational formulation for the dynamics of a semicoherent hetero-interface depends on the physical setting (e.g. different mobilities) of the dislocation and disconnection structure and dynamics, and we will not further discuss the details of them in this paper.

The variational framework presented in this paper focuses on the cases of single array of dislocations and single mode of disconnections,
which is able to capture the natures of microstructure dependent grain boundary motion, and in the meantime, to provide a mathematically tractable theory. More general cases of multiple arrays of dislocations and multi-modes of disconnections can also be included in this variational framework, following the Onsager principle and detailed structure and dynamics of different arrays of dislocations and modes of disconnections.

\appendix

\section{Solution of the ODE in the variational formulation in Sec.~\ref{sec:var}}\label{appendix A}

Eqs.~\eqref{odev}--\eqref{odelambda} for the velocity $\mathbf v$, the dislocation reaction rate $\mathbf B^0_t$ and the Lagrange multiplier $\pmb\lambda$ have the same form:
\begin{equation}
-a\frac{d^2\mathbf u}{ds^2}+\mathbf u=\mathbf f,
\end{equation}
subject to the periodic boundary conditions
\begin{equation}
\mathbf u(l)=\mathbf u(0), \ {\rm and}\ \frac{d\mathbf u}{ds}(l)=\frac{d\mathbf u}{ds}(0),
\end{equation}
Here $a>0$, and $l$ is the perimeter of the grain boundary.

The solution of this problem is
\begin{equation}
\mathbf u(s)=\left(\mathbf D_1-\frac{1}{2\sqrt{a}}\int_0^s\mathbf f(w)e^{-\frac{w}{\sqrt{a}}}dw\right)e^{\frac{s}{\sqrt{a}}}
+\left(\mathbf D_2+\frac{1}{2\sqrt{a}}\int_0^s\mathbf f(w)e^{\frac{w}{\sqrt{a}}}dw\right)e^{-\frac{s}{\sqrt{a}}},
\end{equation}
where
\begin{flalign}
\mathbf D_1= -\frac{1}{2\sqrt{a}(1-e^{\frac{L}{\sqrt{a}}})}\int_0^L e^{\frac{l-s}{\sqrt{a}}}\mathbf f(s) ds,\ \
\mathbf D_2= \frac{1}{2\sqrt{a}(1-e^{-\frac{L}{\sqrt{a}}})}\int_0^L e^{\frac{s-l}{\sqrt{a}}}\mathbf f(s) ds.
\end{flalign}

\bibliographystyle{plain}
\bibliography{rs_ref}

\end{document}